%% file: apssamp.tex
\documentclass[aps,prl,twocolumn,superscriptaddress,reprint]{revtex4-2}
\usepackage{lipsum}
\usepackage{graphicx}% Include figure files
\usepackage{comment}% comment
\usepackage{dcolumn}% Align table columns on decimal point
\usepackage{bm}% bold math
\usepackage{xcolor}
\usepackage{siunitx}
\usepackage{amsmath}
\usepackage[inline]{enumitem}
\begin{document}

\title{Applying machine learning to determine impact parameter in nuclear physics experiments}% Force line breaks with 

\include{authors}

\date{\today}% It is always \today, today,

\begin{abstract}
Machine Learning (ML) algorithms have been demonstrated to be capable of predicting impact parameter in heavy-ion collisions from transport model simulation events with perfect detector response. We extend the scope of ML application to experimental data by incorporating realistic detector response of the S$\pi$RIT Time Projection Chamber into the heavy-ion simulation events generated from the UrQMD model to resemble experimental data.  At \SI{3}{fm}, the predicted impact parameter is \SI{2.8}{fm} if simulation events with perfect detector is used for training and testing; \SI{2.4}{fm} if detector response is included in the training and testing, and \SI{5.8}{fm} if ML algorithms trained with perfect detector is applied to testing data that has included detector response. The last result is not acceptable illustrating the importance of including the detector response in developing the ML training algorithm.  We also test the model dependence by applying the algorithms trained on UrQMD model to simulated events from four different transport models as well as using different input parameters on UrQMD model. %The combined effect from detector response and model dependence worsens the ML determination accuracy and resolution by $\approx 10\%$. 
Using data from Sn+Sn collisions at E/A=\SI{270}{MeV}, the ML determined impact parameters agree well with the experimentally determined impact parameter using multiplicities, except in the very central and very peripheral regions. ML selects central collision events better and allows impact parameters determination beyond the sharp cutoff limit imposed by experimental methods. 

\begin{comment}
\begin{description}
\item[Usage]
Secondary publications and information retrieval purposes.
\item[PACS numbers]
May be entered using the \verb+\pacs{#1}+ command.
\item[Structure]
You may use the \texttt{description} environment to structure your abstract;
use the optional argument of the \verb+\item+ command to give the category of each item. 
\end{description}
\end{comment}
\end{abstract}

\maketitle

%\tableofcontents

\section{\label{sec:intro} Introduction}

With the availability of inexpensive computing power and open-source Machine learning (ML) algorithms, the application of ML to daily and scientific problems has advanced significantly in recent years. One attractive feature of ML is the ability of these algorithms to ``learn" or ``generalize" from training data. For instance, to train for optical character recognition, the training data will consist of both the images of handwritten characters and their corresponding character label~\cite{Deng12}. Machine learning algorithms will figure out the high dimensional correlation between the image and character label on its own without being programmed to explicitly look for a particular pattern, such as matching certain characteristic strokes or applying certain transformations ~\cite{Gader95}. Once trained, the algorithm should be able to predict the underlying character from an image of a character outside of the training set.

Increasingly, ML has been applied to science. After all, science is to discover the fundamental physics from regular patterns in nature such as the motion of the planets and constellation of the stars. One is also hopeful that ML will spot the high dimensional correlations in nature not perceived by human due to our physical and intellectual limitations.

 Nuclear collision experiments designed to study the bulk properties of nuclear matter depends on the initial proximity of the projectile and target. For example, one uses head-on collisions to study fusion but peripheral collisions to study elastic scattering. Thus the determination of the centrality of a collision, quantified by impact parameter ($b$) which is defined as the closest distance between a target and a projectile, is central to extract physics insights from the experimental results. This is especially the case in the study of the nuclear equation of state using heavy-ion collisions~\cite{estee2021,Jhang21} which are of interest not only in understanding the existence of nuclei but also in understanding the properties of neutron stars. Depending on the energy and impact parameters, the density and temperature of the nuclear matter created in the fleeting moment when the projectile and target overlap can differ by orders of magnitude.

Machine Learning algorithms have long demonstrated their potentials in determining impact parameters in heavy-ion collisions~\cite{Bass94, Bass96, De_Sanctis_2008, Christophe1995, Oma19}. Recently, the ML algorithms of CNN and LightGBM have been used successfully to extract the impact parameters from simulated Au+Au collisions~\cite{WANG2020} and very recently Sn+Sn collisions using transport models~\cite{li2021application}. However, before applying the ML algorithms to experimental data, the simulated events must undergo the same conditions experienced by real particles before detection and recorded as data. In this paper, we will show the results of applying the ML algorithm to the real data from the $^{132}$Sn+$^{124}$Sn collisions at \SI{270}{AMeV} taken by the S$\pi$RIT Time Projection Chamber to determine the impact parameters event by event. 

\section{\label{sec:TPC} S$\pi$RIT Time Projection Chamber}
In simulations, particles are emitted at all angles with all energies. However, in experiments, there are physical limitations due to the placements of detectors and the design of the experimental setup which is generally optimized for specific goals of the experiment to detect certain particles of interest. For example, a complete $4\pi$ geometrical coverage is impossible, because detectors cannot be placed in space without mechanical support which would block some of the phase space. Both the detector and the electronics that process the detector signals have a limited dynamic range so only certain particles with a limited energy range are measured. Depending on the gain and signal to noise requirements, the lightest or the heaviest particles could be excluded from detection or not optimally detected ~\cite{Estee19}. Such detector limitations must be taken into account in the training of data using the machine learning algorithm. 

The data discussed in this paper has been acquired by the S$\pi$RIT collaboration. The main detector in the experiment is the SAMURAI-pion-Reconstruction and Ion-Tracker (S$\pi$RIT) Time Projection Chamber (TPC) which is a large rectangular gas detector designed to study the nuclear equation of state around twice the normal nuclear matter density $\rho_0 \approx \SI{2.6e14}{\gram\per\centi\metre\cubed}$ or $\SI{0.16}{\per\femto\metre\cubed}$. In a recent experiment, a radioactive $^{132}$Sn beam accelerated to \SI{270}{AMeV} is impinged on an isotopically enriched $^{124}$Sn targets to create a neutron-rich system to study the properties of neutron stars and dynamics of nuclear collisions. A high-density region, up to 2$\rho_0$, can be created in the central collisions of heavy nuclei. The properties of this dense region are deduced by studying the emitted particles detected by the S$\pi$RIT TPC installed inside the SAMURAI magnet. The design, construction, operation and performance of the TPC as well as the complicated data analysis techniques have been published in several papers~\cite{Barney20, Kobayashi13,Tsangc20,Estee19, Iso18, Lee20, Las17, Jhang21, Shane15}. In order to selectively enhance the central collision events, auxiliary or trigger detectors are used to veto out peripheral events. Due to its complexity and corrections needed to apply to the raw data, the data acquired by the S$\pi$RIT experiment is ideal to study the effect of the detector response in machine learning.

For comparison with the machine learning results, we describe here how the impact parameters are determined experimentally using the number of charged particle detected in an event, referred to as multiplicity, $M$; an observable that the TPC is especially efficient in detecting. $M$ is monotonically proportional to $b$ assuming geometric cross-section $d\sigma = 2\pi bdb$ using the following formula~\cite{Barney19},
\begin{equation}
b(M_C) = b_\text{max}\frac{N_{M\geq M_C}}{N_\text{total}},
\label{eq:trad}
\end{equation}
where, $N_{M\geq M_C}$ is the number of observed events with M $\geq M_C$, $N_\text{total}$ is the total number of observed events, and $b_\text{max}$ is the maximum impact parameter which can be determined by measuring the reaction cross-section with the trigger bias. One can generalize this method to other observables as long as the observable is monotonically proportion to $b$. For clarity, this method will be referred to as Multiplicity in all the figures and text below. 

\section{\label{sec:lightGBM} Machine Learning Algorithms}

The impact parameter determination is a regression problem that can be solved by machine learning once training data is obtained. Based on the detailed study in the Ref.~\cite{li2021application}, we use the LightGBM to extract impact parameters first from simulation data obtained from the Ultra-relativistic Quantum Molecular Dynamics (UrQMD) model and then from experimental data. %Our ML algorithm of choice is LightGBM as its performance is similar to CNN but much less CPU intensive. Details of the ML algorithm including the ultra-relativistic quantum molecular dynamics (UrQMD) model, the transport model used to generate events for training are described in the companion paper [YJWang2021]. 
We do not expect the general conclusion of this work to change significantly if another ML algorithm or another transport model is used.

Seven observables~\cite{li2021application} found to correlate with impact parameters from model and experimental studies are chosen as inputs or features in the training: \begin{enumerate*}[label=(\roman*)]  \item Total multiplicity of charged particles. \item Transverse kinetic energy of hydrogen and helium isotopes. \item Ratio of total transverse-to-longitudinal kinetic energy. \item Total number of hydrogen and helium isotopes. \item Averaged transverse momentum of hydrogen and helium isotopes. \item Number of free protons at mid-rapidity $|y_{z}/y_{beam}|\leq0.5$. \item Averaged transverse momentum of free protons at mid-rapidity $|y_{z}/y_{beam}|\leq0.5$.\end{enumerate*} 
 
To quantify the performance of algorithm, we use bias (mean deviation) and standard deviation (S.D.) of the predicted impact parameter. The bias and S.D. quantify accuracy and precision, respectively, and are defined as:
\begin{equation}
\begin{split}
\text{Bias}~(b^{\text{pred}}) &= \overline{b^{\text{pred}} - b^\text{true}} \\
\text{S.D.} = \sqrt{\text{Var}~(b^{\text{pred}})} &=  \sqrt{\overline{(b^{\text{pred}} - \overline{b^{\text{pred}}})^2}}.
\end{split}
\label{eq:db}
\end{equation} 
Here, $b^\text{true}$ is the true impact parameter used in event generation and $b^\text{pred}$ is the mean predicted impact parameter from the LightGBM. The averaging is done over each event.  

The results of the current work are based on the LightGBM algorithm trained on 135,000 UrQMD simulated events for the reactions of $^{132}$Sn+$^{124}$Sn at \SI{270}{AMeV}. For the training data, the parameter set known as ``SM-F'' is used as the input parameters to UrQMD model. In SM-F, the incompressibility $K_0$ is set to \SI{200}{MeV} and the nucleon-nucleon (NN) elastic scattering cross-section in free space is used as the in-medium NN cross-section. Another input parameter set SM-I that employs a different in-medium NN cross-section~\cite{li2021application} will also be tested together with other transport models.

\section{Results and discussion}

In this section, we will apply the response of the S$\pi$RIT time-projection chamber (TPC)~\cite{Barney19} to test the generalizability of ML algorithms and to examine the effect of detector response to the impact parameter determination. 

In the  S$\pi$RIT experiment, both the efficiency and fragment momentum resolution depend on the emission angle as well as the environment of the TPC at the time of detection. As a result, one cannot approximate the S$\pi$RIT experimental response as a simple filter that removes data outside of the detection region which is a common technique~\cite{Reisdorf10}. Instead, we develop a Monte Carlo package based on GEANT4~\cite{Agostinelli03} that converts events generated from transport models into detector electronic signals by incorporating the physics of detector gas ionization and detector digitization. The resulting signals that resemble data will be analyzed in the same way as real experimental data using the same data analysis package SpiRITROOT~\cite{Lee20, Estee19, Iso18} specially developed for the S$\pi$RIT TPC. The UrQMD events reconstructed after running through SpiRITROOT will then be used for training and testing the applicability of impact parameter algorithms in a realistic setting.

\begin{figure}
\includegraphics[width=1\linewidth]{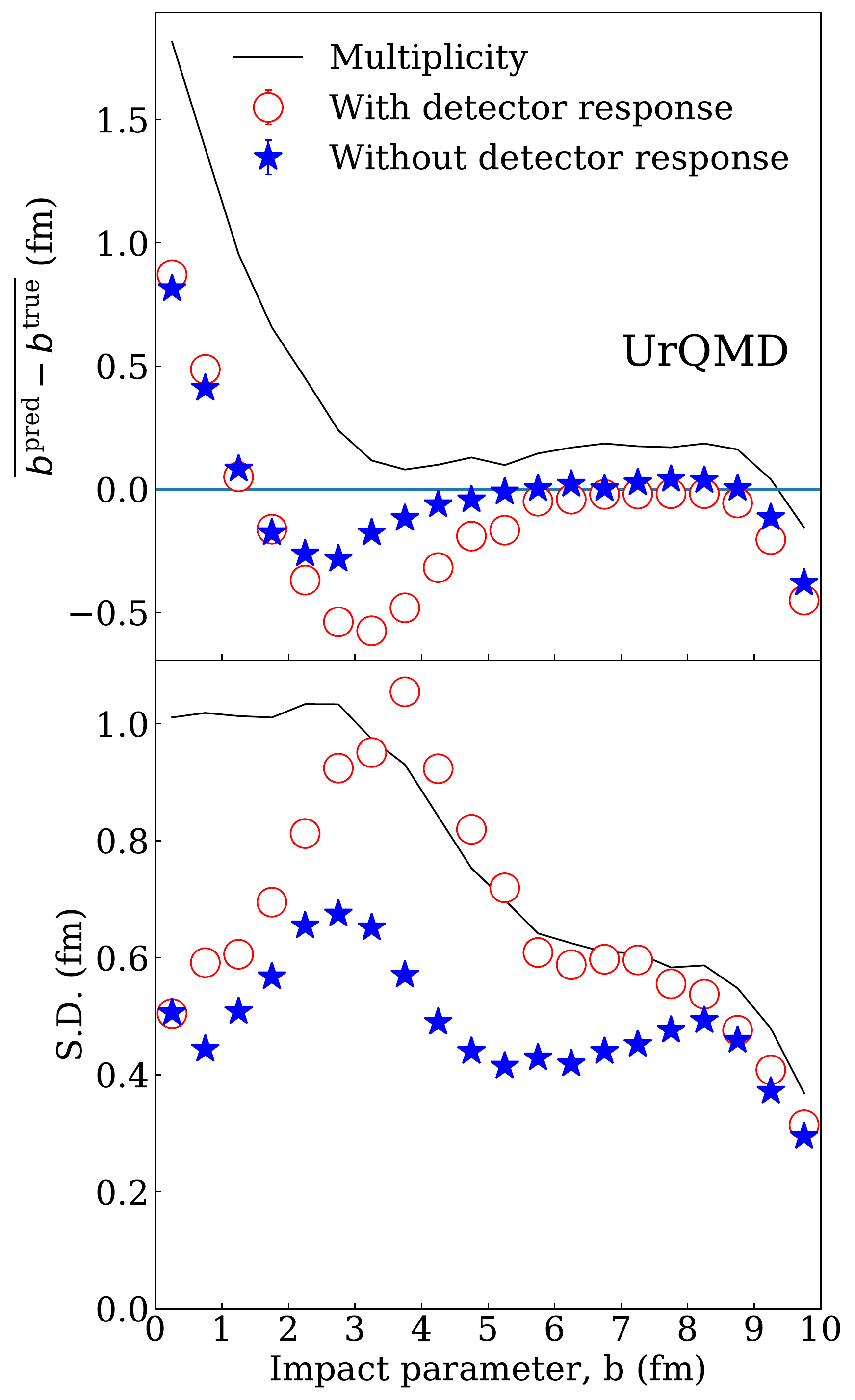}
\caption{Impact parameter dependence of bias (upper panel) and S.D. (lower panel) predicted by LightGBM without detector response (solid stars) and with detector response (open circles). }
\label{fig:predVsTruth}
\end{figure}

Fig.~\ref{fig:predVsTruth} shows bias (top panel) and S.D. (bottom panel) as a function of impact parameter ($b^{true}$). The LightGBM is used to train and test on two data sets: one includes the response of the S$\pi$RIT experiment (open red circles) and one without (blue solid stars). %The discussion of the black line is delayed to later paragraphs. 
Both the training and the testing data sets use the same UrQMD input parameter set of SM-F. As expected both bias and S.D. worsens with the inclusion of detector response especially in the mid-peripheral regions. Around $b$=\SI{3}{fm}, both bias and S.D. worsens by a factor of 2 when detector response is included. The worsening in bias and S.D. even when detector response is not included could be related to the physics of transport models. In central collisions (small $b$), nucleon-nucleon scatterings dominate in the collision dynamics while in peripheral collisions (large $b$), the mean field dominates. In the mid-central or mid-peripheral regions ($b$=3-\SI{5}{fm}), accurate treatment of both the mean-field and collisions are very important but transport models may fall short. 

The black curves represent ``Multiplicity" results obtained using Eq.~\eqref{eq:db}. In general, the Multiplicity algorithm performs worse than ML especially, for central collisions. Experimentally, we also see that ML selects central collision events better as discussed below.  % results. For the bias, results from Multiplicity and LightGBM are comparable with the Multiplicity method being slightly worse beyond $b>5 fm$. For S.D., the similarity starts with $b>3 fm$. 

For perfect determination of $b$, both bias and S.D. should be zero or close to it. That happens for the bias only in the range of $b$=5-\SI{8}{fm}.  Over this region, the detector effects are minimal. S.D. never approaches zero over the range of $b$ we investigate.  The worsening of both bias and S.D. around $b\approx\SI{0}{fm}$ and $b\approx\SI{10}{fm}$ have been observed with other ML algorithms. This could be 
due to the inability of the LightGBM and the multiplicity algorithms to predict accurately near the boundaries of a data set. 

%In the following discussion, we focus on the region between $b=2-\SI{5}{fm}$ where the effect of detector response is most significant. 
%In general, the increases in bias due to detector response are   modest, probably because the uncertainties in bias from ML is larger than that from detector response. 

%The effect of detector response on S.D. is much larger ($>50\%$) except for very large $b>\SI{8}{fm}$. Again, S.D. values in the region $b=1.5-\SI{5.5}{fm}$ increase. Including the experimental response significantly worsens both the accuracy and precision around \SI{3.5}{fm} by nearly a factor of 2. 

% moved table forward to force it to appear sooner
\begin{table*}[htbp]
\caption{Statistical properties of $b^\text{pred}$ on simulated events from various transport models. Simulated data from UrQMD/SM-F input parameter set are used for training. The bias values are plotted as absolute numbers. All values are in unit of fm.}
\begin{tabular}{p{0.12\linewidth}>{\centering}p{0.05\linewidth}>{\centering}p{0.12\linewidth}>{\centering}p{0.12\linewidth}>{\centering}p{0.12\linewidth}>{\centering\arraybackslash}p{0.12\linewidth}>{\centering\arraybackslash}p{0.12\linewidth}>{\centering\arraybackslash}p{0.12\linewidth}}
\hline
Model&AMD&dcQMD&ImQMD&IQMD&Average&UrQMD/SM-F&UrQMD/SM-I\\
\hline
\\[-1ex]
\multicolumn{2}{l}{Perfect detector} & & & & & & \\[1ex]
\multicolumn{1}{l}{$\overline{b^\text{pred}}$} & 4.09 & 2.84 & 3.29 & 3.19 & 3.35 & 2.77 & 3.20\\
\multicolumn{1}{l}{S.D.} & 0.68 & 1.00 & 0.74 &0.88 &   0.83 & 0.66 & 0.94\\
\multicolumn{1}{l}{$|\text{Bias}|$} & 1.09 & 0.16 & 0.29 &0.43 &   0.49 & 0.23 & 0.20\\
\hline
\\[-1ex]
\multicolumn{2}{l}{With realistic detector response} & & & & & \\[1ex]
\multicolumn{1}{l}{$\overline{b^\text{pred}}$} & 4.06 & 3.77 & 3.22 & 2.66 & 3.43 & 2.44 & 2.96\\
\multicolumn{1}{l}{S.D.} & 0.91 & 1.22 & 1.02 & 1.03 & 1.04 & 0.94 & 1.05\\
\multicolumn{1}{l}{$|\text{Bias}|$} & 1.06 & 0.77 & 0.22 &0.34 & 0.60 & 0.56 & 0.04\\
\hline
\\[-1ex]
\multicolumn{8}{l}{Trained with perfect detector, applied to simulation with detector response}\\[1ex]
\multicolumn{1}{l}{$\overline{b^\text{pred}}$} & 6.45 & 6.45 & 6.25 & 6.04 & 6.30 & 5.87 & 5.69\\
\multicolumn{1}{l}{S.D.} & 0.44 & 0.46 & 0.44 & 0.46 & 0.45 & 0.62 & 0.46\\
\multicolumn{1}{l}{$|\text{Bias}|$} & 3.45 & 3.45 & 3.25 & 3.04 & 3.30 & 2.87 & 2.69\\
\hline
\label{tab:modelDep}
\end{tabular}
\end{table*}

%A comparison of Multiplicity and LightGBM is performed in Fig.~\ref{fig:predVsTruth} where the black line corresponds to bias and S.D. of Multiplicity inference when applied to simulated data with detector response. Multiplicity exhibit the same worsening of bias below \SI{3}{fm} as ML algorithms. This is caused by the fact that $b^\text{pred}$ is limited to positive values which restricts how negative $b^\text{pred} - b^\text{true}$ can be. The negative values is insufficient to average out the positive values. 

%It is observed that LightGBM shows smaller value S.D. for central events with $b < \SI{3}{fm}$, but their performances are similar in mid central to peripheral regions. However, when multiplicity inference is applied to experimental data, it cannot work on peripheral regions beyond cross-section measurement of \SI{7.5}{fm} given by Ref.~\cite{Barney19}. The S$\pi$RIT experiment is designed to study the equation of state from central collisions. Peripheral collisions are disproportionately rejected in the experimental data by the triggers, an effect that is not simulated in the S$\pi$RITROOT package. As a result LightGBM has an advantage in predicting impact parameter of peripheral events.

To verify that application of the LightGBM algorithm trained on the UrQMD simulations is not restricted to only simulations from the UrQMD model and can be generalized to experimental data, we test the algorithm using simulations from four different transport models, Antisymmetrized Molecular Dynamics (AMD) model~\cite{Ike16, Ike20} plus three different families of Quantum Molecular Dynamics (QMD) models, dcQMD~\cite{Coz16,Coz17}, IQMD~\cite{hartnack1998,le2016} and ImQMD~\cite{ZHANG14}. All these models, including UrQMD, use different techniques and approaches to simulate the nucleus-nucleus collisions.  All of them have had various success in describing heavy ion collision data. Refs.~\cite{Xu16, Zhang18} detail some of the major differences among these models and their performance in heavy-ion collision simulations. In the simulations described here, default physics input parameters for each code are used. This allows us to not only gauge the discrepancy caused by different model assumptions, but also by uncertainty in input parameter values.
In addition to these four different models, we also include a different input parameter set for the UrQMD model. The results labeled as UrQMD/SM-I can be considered as a different model. 

To quantify the effects of how ML is applied to data, we perform three types of tests at $b=\SI{3}{fm}$. The results are listed in the top, middle, and bottom sections of Table I. The top section contains results from the perfect detector (i.e., without the inclusion of detector response to simulated events) both for training and testing. The middle section contains results from including the detector response for both training and testing. Finally, in the bottom section, we apply the ML algorithm trained with perfect detectors to testing events that include detector response. The last option gives the largest deviation of $b^{true}$ by predicting the mean $b^{pred}$ as nearly \SI{6}{fm}. Such results and, therefore, the algorithm not including detector response in the training is unacceptable and would not be discussed any further.

About 5000 events at \SI{3}{fm} are generated from each code. LightGBM trained with UrQMD/SM-F data is tasked with predicting the impact parameter of these events. Reference for best performance is established by training and testing events directly from UrQMD without detector response. The values are listed under the label UrQMD/SM-F in the top section of Table~\ref{tab:modelDep}.  %The bias is represented with absolute values so that positive and negative values do not cancel each other when an average is calculated. The S.D. increases by nearly $\sim 50\%$, from $\sim\SI{0.66}{fm}$ to \SI{0.94}{fm} when the training and testing data sets are different even if the simulating model is the same.

The bias and the corresponding S.D. values for AMD, dcQMD, ImQMD, and IQMD and their average are listed in Table~\ref{tab:modelDep}. Here, the AMD has the largest bias (\SI{1.09}{fm}) reflecting the very different approaches used in simulating HIC in AMD and other QMD-type models. As expected since these transport models were not used to train the events, both the bias and S.D. are larger than  those values listed under UrQMD/SM-F column in Table~\ref{tab:modelDep}. Except for AMD, the bias and S.D. from different transport models are similar to the results of UrQMD/SM-I where the training and testing data use different input parameter sets. For AMD, while the accuracy worsens, the S.D. values are similar to the reference of UrQMD/SM-F. 

If the average performance of the different models is compared to the reference, S.D. increases by 20\%. The UrQMD/SM-F underpredicts while the other models overpredict $b$.  As expected, including the detector response worsens the bias and S.D. for all models. Assuming that the data could be described by the average of the models, then one could expect that the ML algorithm could determine $b$ to within \SI{0.6}{fm} with S.D. of \SI{1}{fm} from experimental data. % i.e. at $b$=\SI{3}{fm}, the uncertainty in $b$ determination is about \SI{1.2}{fm}. %The effect of detector response is much less outside this region.

\begin{figure}
\includegraphics[width=\linewidth]{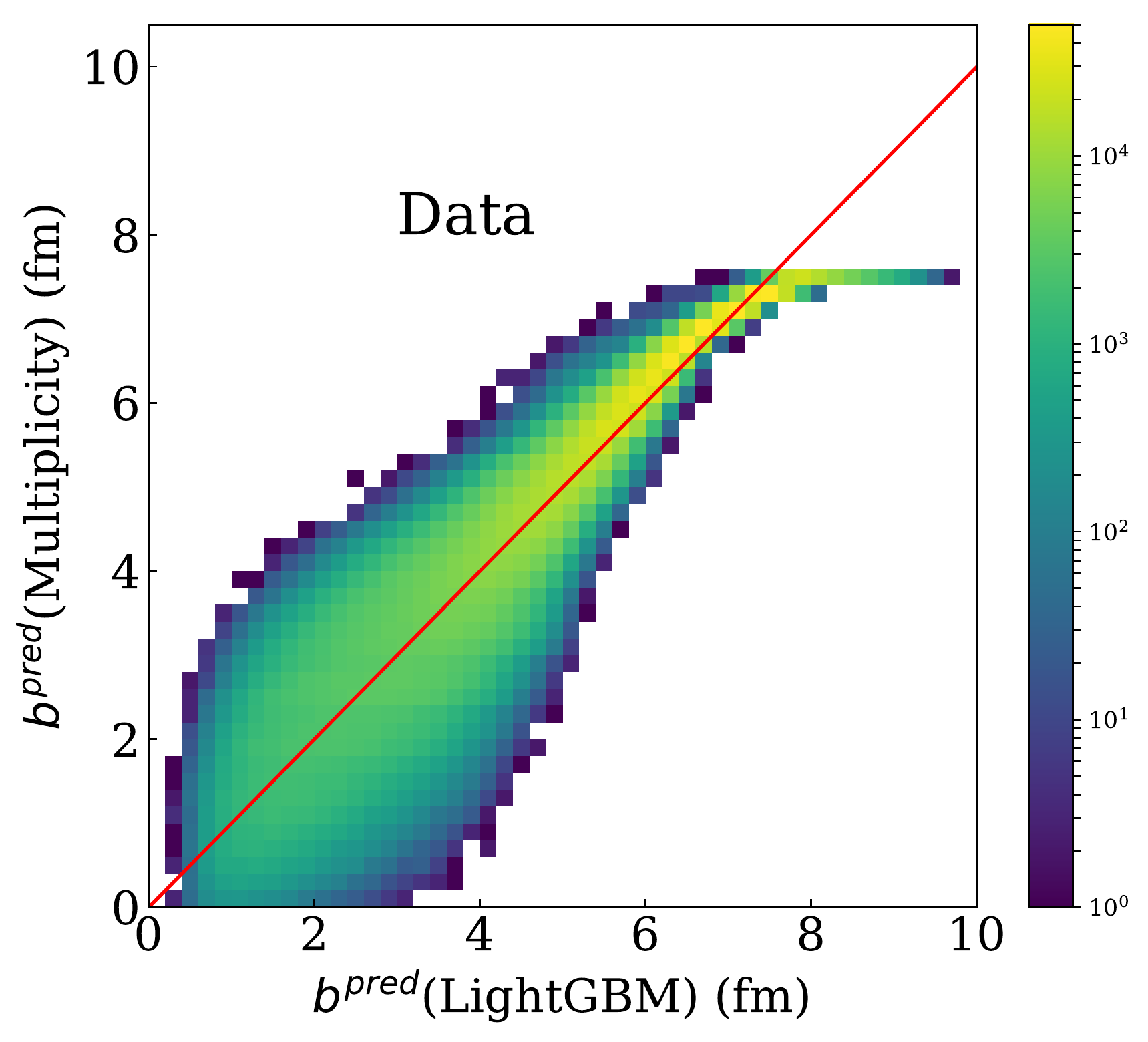}
\caption{Here, the $b^\text{pred}$ values from the LightGBM is plotted against that from multiplicity. Color represent number of counts in each bin. The red diagonal line shows the expected correlation if impact parameters are determined perfectly.}
\label{fig:bMLVsMult}
\end{figure}

Finally, we apply the ML algorithm to the data. Fig.~\ref{fig:bMLVsMult} plots the correlations between $b^\text{pred}$ from LightGBM and $b^\text{pred}$ from Multiplicity of eq.~\eqref{eq:trad}. Generally, they are in agreement as evidenced by the overall diagonal distribution. Experimental cross-section measurements sets $b_\text{max}$ from multiplicity to be \SI{7.5}{fm} while $b^\text{pred}$ from the LightGBM extends beyond the sharp cut off limit resulting in a horizontal tail at $\SI{7.5}{fm}$. 

\begin{figure}
\includegraphics[width=\linewidth]{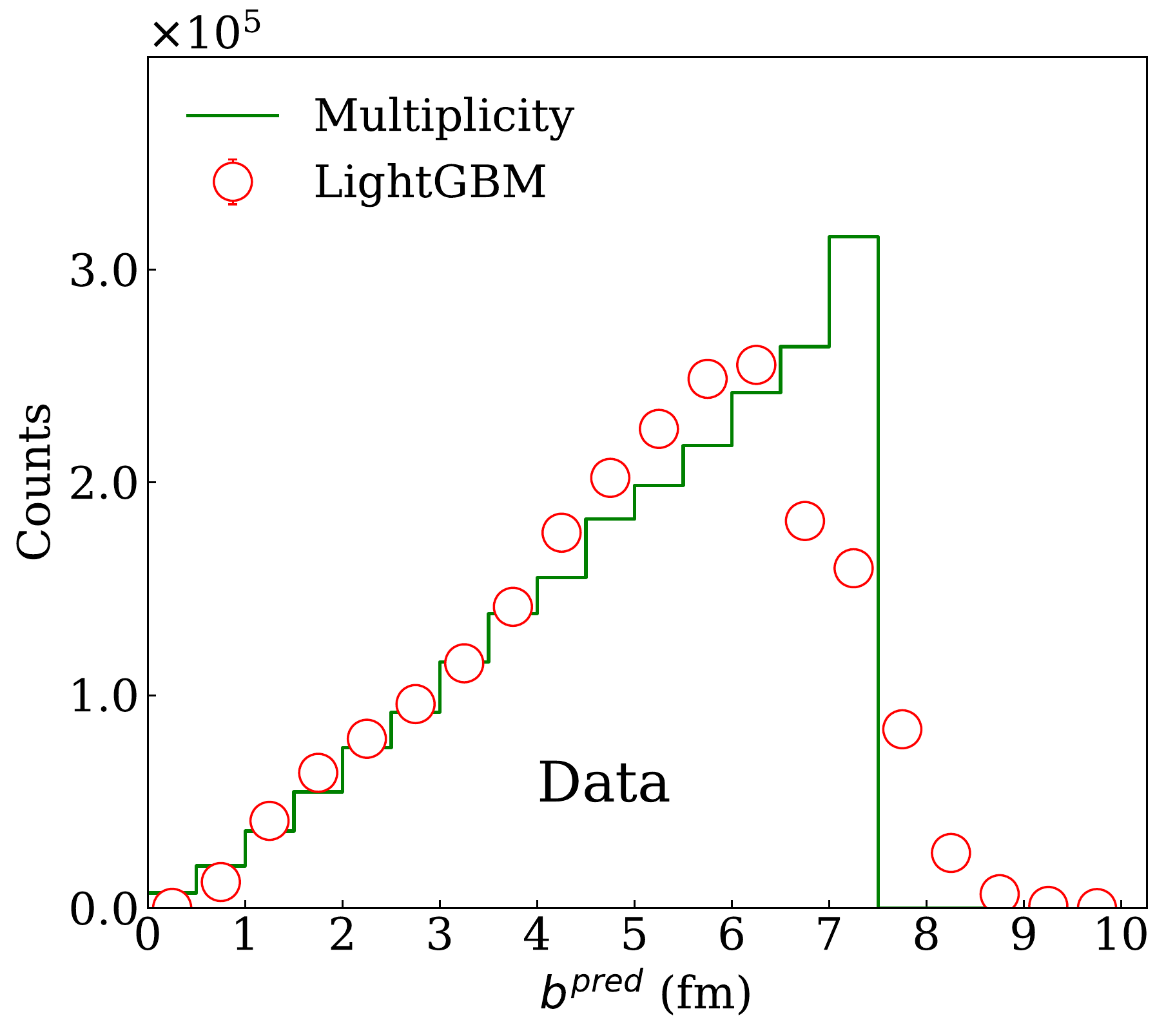}
\caption{Distribution of $b^\text{pred}$ made with the LightGBM (open symbols) and multiplicity (histogram). The histogram represents sharp cut off model with $b_\text{max}=\SI{7.5}{fm}$.}
\label{fig:bdist}
\end{figure}

The histogram in Fig.~\ref{fig:bdist} shows the experimental impact parameter distributions from the sharp cut off model of eq.~\eqref{eq:trad}. The impact parameter distribution predicted by the LightGBM (open symbols) exhibits a tail that extends $b^\text{pred}$ beyond \SI{7.5}{fm}. It resembles smearing of the experimental multiplicity distribution which is consistent with the expectation that the experimental data should contain a range of impact parameters that would extend beyond $b_\text{max}$. In addition, one would expect the sharp cutoff model multiplicity distribution should always be equal to or higher than the realistic multiplicity distributions. Fig.~\ref{fig:bdist} shows that from 4 to \SI{6.5}{fm}, there are slightly more events from LightGBM than from Multiplicity analysis. This apparent discrepancy is not understood. It could be that, not all the detector response has been accurately reproduced. It could also be that the UrQMD is not describing the experimental data accurately enough in this region as is also evidenced by the worsening of the accuracy and broadening of S.D. in Fig.~\ref{fig:predVsTruth}. Nonetheless, the effects are small.

\begin{comment}

As for pion yield, it is expected to increase with decreasing impact parameter. The high production threshold of pion means they are only created in dense region~\cite{Ikeno16, Li02}. Ref.~\cite{Jhang21} detailed the analysis procedure required to extract reliable $\pi^-$ yield from S$\pi$RIT, and it is plotted against $b^\text{pred}$ in Fig.~\ref{fig:pionyield}. The fact that the distribution for LightGBM continues to decrease beyond the sharp-cut off of \SI{7.52}{fm} indicates its applicability on peripheral events. While LightGBM and multiplicity performs similarly at low impact parameter, ERAT performance is significantly worse. Although each observable individually only indirectly hints at the performance of LightGBM, the fact that it performs well simultaneously on both is a strong indication of its effectiveness. 

\begin{figure}
\vspace*{0.5cm}
\includegraphics[width=\linewidth]{pi_yield.png}
\caption{$\pi^-$ yield as a function of $b^\text{pred}$. The red line corresponds to }
\label{fig:pionyield}
\end{figure}

\end{comment}

Unlike events from transport models, we do not have the true value of impact parameter from experimental data so we cannot evaluate the accuracy of $b^\text{pred}$  values. Fig.~\ref{fig:bdist} suggests that the LightGBM algorithm determines the impact parameter for peripheral events more accurately as it does not have the sharp cutoff limit and the impact parameter smearing occurs naturally. To evaluate the performance at central collisions, we use observables whose qualitative behavior with impact parameter is known. 

One such observable is the reaction plane resolution $\langle\cos(\Phi_{M}-\Phi_{R})\rangle$\cite{poskanzer1998methods}. Here $\Phi_{M}$ and $\Phi_{R}$ are the measured and the real azimuthal angle of the reaction plane, respectively. The reaction plane should vanish  as $b$ approaches zero due to azimuthal symmetry. In a perfect head-on collision ($b = \SI{0}{fm}$), the fragment emission is isotropic and $\Phi_{M}$ is reduced to a random number between $0 - 2\pi$, which makes the average of cosine zero.

\begin{figure}
\vspace*{0.5cm}
\includegraphics[width=\linewidth]{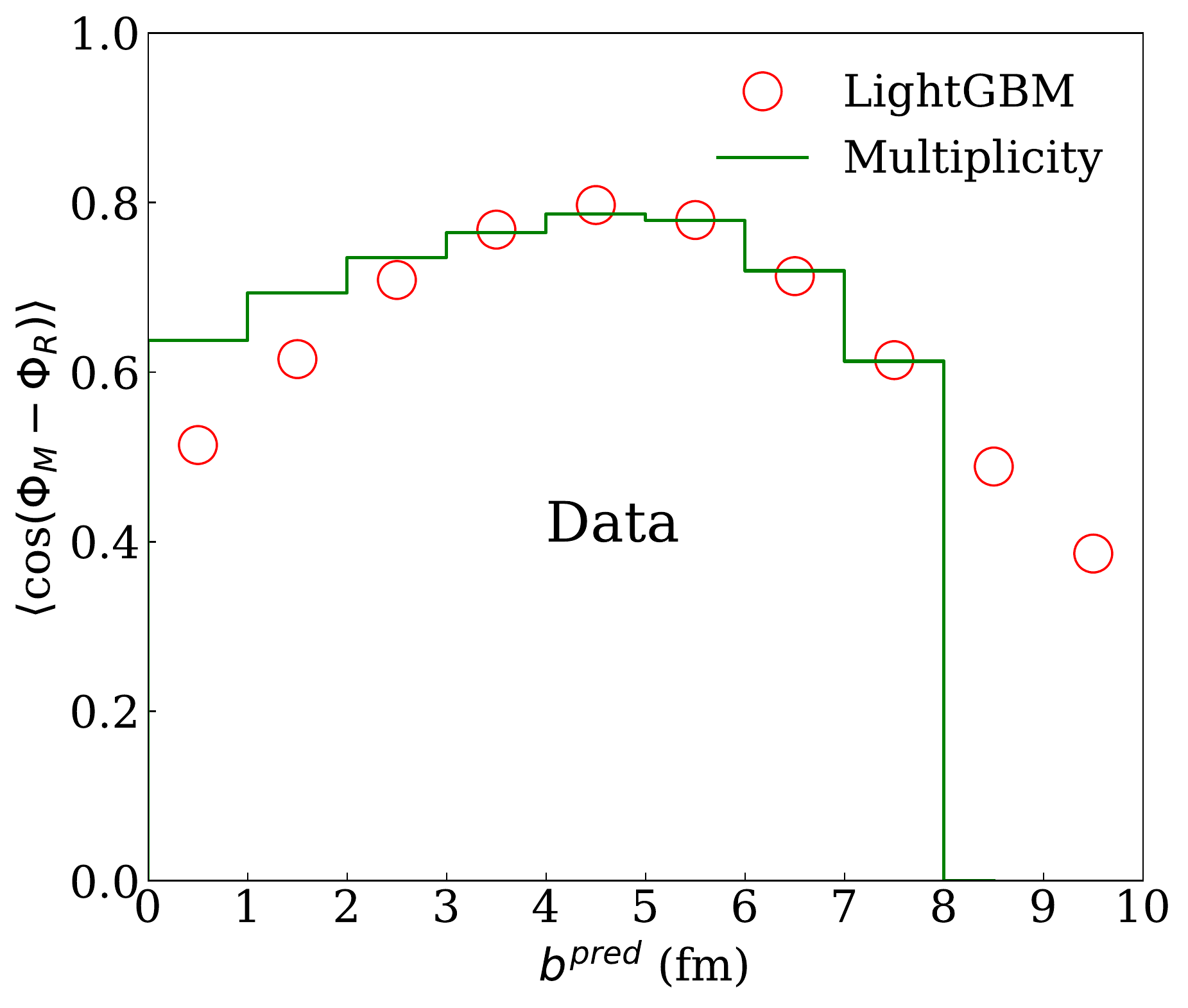}
\caption{The reaction plane angle resolution, $\langle\cos(\Phi_{M}-\Phi_{R})\rangle$ is plotted against $b^\text{pred}$. The predictions are made with Multiplicity (green histogram) and LightGBM (red open circle).}
\label{fig:RPResolution}
\end{figure}

As shown in Fig.~\ref{fig:RPResolution}, the reaction plane resolution $\langle\cos(\Phi_{M}-\Phi_{R})\rangle$ decreases with $b^\text{pred}$. However, at $b^\text{pred} < \SI{3}{fm}$ the reaction plane resolution is closer to zero if the central event selections are made with LightGBM. This finding supports the assertion that events selected by LightGBM are more central than the corresponding events selected by multiplicity,  although neither intercepts the y-axis at zero.

\section{Summary}
In summary, the scope of the machine learning algorithm i.e., the LightGBM is extended to determine impact parameter for the experimental data. Previously, the algorithm demonstrated its capability to determine impact parameter from the UrQMD simulated events i.e. using a perfect detector. Here, we trained the ML algorithm to analyze the UrQMD events incorporated with realistic detector response of the S$\pi$RIT Time Projection Chamber to resemble experimental data. The model dependence of ML algorithm is also tested by using events generated from 4 different transport models as well as using a different input parameter set on the trained model at $\SI{3}{fm}$ where the performance of the UrQMD is the worst. As long as the training and testing data are consistent in incorporating the detector response, the accuracy of the impact parameter determination is less than \SI{1}{fm}. We used the ML algorithm to determine impact parameter for the experimental data of Sn+Sn collisions at E/A=\SI{270}{MeV} detected by the S$\pi$RIT Time Projection Chamber. Except for the very central and very peripheral regions, the obtained impact parameters agree with the ones experimentally determined using the multiplicity method. Furthermore, the ML algorithm is able to better select the central collision events and is also able to predict events beyond \SI{7.5}{fm}, which multiplicity inference or similar methods cannot do due to the sharp cutoff approximations to account for trigger bias in the experimental setup.        

% When applied to experimental data, it shows good agreement with inference from multiplicity using equation~\eqref{eq:trad}. 

%When applied to experimental data, analysis with impact parameters determined by the ML algorithm shows a smaller proton $v_1$ than the corresponding values with impact parameter selected by traditional multiplicity, indicating that the ML algorithm is better suited for selecting central events. In addition, without the sharp cut off at $b_\text{max}$ ML allows $v_1$ to be determined beyond \SI{7.5}{fm} due to smearing of the impact parameters, resembling real data.
%and larger $\pi^-$ yield than by ERAT 
\section*{Acknowledgement}

This work was supported by the U.S. Department of Energy, USA under Grant Nos. DE-SC0014530, DE-NA0003908, U.S. National Science Foundation Grant No. PHY-1565546, the National Science Foundation of China Nos. U2032145, 11875125, and 12047568,
and the National Key Research and Development Program of China under Grant No. 2020YFE0202002, and the ``Ten Thousand Talent Program" of Zhejiang province (No. 2018R52017), the Japanese MEXT, Japan KAKENHI (Grant-in-Aid for Scientific Research on Innovative Areas) grant No. 24105004, the National Research Foundation of Korea under grant Nos. 2016K1A3A7A09005578, 2018R1A5A1025563,2013M7A1A1075764,and the Romanian Ministry of
Education and Research through Contract No. PN 19 06 01 01/2019-2022. Computing resources were provided by the NSCL and the Institute for Cyber-Enabled Research at Michigan State University.

\input{output.bbl}

\end{document}

%% file: authors.tex


\author{C.Y.~Tsang}%~(\CJKfamily{bsmi}{曾}}
\affiliation{National Superconducting Cyclotron Laboratory, Michigan State University, East Lansing, Michigan 48824, USA}
\affiliation{Department of Physics, Michigan State University, East Lansing, Michigan 48824, USA}

\author{Yongjia Wang}
\affiliation{School of Science, Huzhou University, Huzhou 313000, China}

\author{M.B.~Tsang}%~(\CJKfamily{bsmi}{曾敏兒})}
%\altaffiliation{Corresponding author}
%\email{tsang@nscl.msu.edu}
\affiliation{National Superconducting Cyclotron Laboratory, Michigan State University, East Lansing, Michigan 48824, USA}
\affiliation{Department of Physics, Michigan State University, East Lansing, Michigan 48824, USA}

\author{J.~Estee}
%\altaffiliation{Corresponding author}
%\email{esteejus@mit.edu}
\affiliation{National Superconducting Cyclotron Laboratory, Michigan State University, East Lansing, Michigan 48824, USA}
\affiliation{Department of Physics, Michigan State University, East Lansing, Michigan 48824, USA}

\author{T.~Isobe}%~(\CJKfamily{min}{磯部忠昭})}
%\altaffiliation{Corresponding author}
%\email{isobe@riken.jp}
\affiliation{RIKEN Nishina Center, Hirosawa 2-1, Wako, Saitama 351-0198, Japan}

\author{M.~Kaneko}%~(\CJKfamily{min}{金子雅紀})}
\affiliation{RIKEN Nishina Center, Hirosawa 2-1, Wako, Saitama 351-0198, Japan}
\affiliation{Department of Physics, Kyoto University, Kita-shirakawa, Kyoto 606-8502, Japan}

\author{M.~Kurata-Nishimura}%~(\CJKfamily{min}{倉田-西村美月})}
\affiliation{RIKEN Nishina Center, Hirosawa 2-1, Wako, Saitama 351-0198, Japan}

\author{J.W.~Lee}%~(\CJKfamily{mj}{이정우})}
\affiliation{Department of Physics, Korea University, Seoul 02841, Republic of Korea}

\author{Fupeng Li}
\affiliation{School of Science, Huzhou University, Huzhou 313000, China}
\affiliation{College of Science, Zhejiang University of Technology, Hangzhou 310014, China}
\author{Qingfeng Li}
\affiliation{School of Science, Huzhou University, Huzhou 313000, China}
\affiliation{Institute of Modern Physics, Chinese Academy of Sciences, Lanzhou 730000, China}

\author{W.G.~Lynch}%~(\CJKfamily{bsmi}{連致標})}
%\altaffiliation{Corresponding author}
%\email{lynch@nscl.msu.edu}
\affiliation{National Superconducting Cyclotron Laboratory, Michigan State University, East Lansing, Michigan 48824, USA}
\affiliation{Department of Physics, Michigan State University, East Lansing, Michigan 48824, USA}

\author{T.~Murakami}%~(\CJKfamily{min}{村上哲也})}
%\altaffiliation{Corresponding author}
%\email{murakami.tetsuya.3e@kyoto-u.ac.jp}
\affiliation{RIKEN Nishina Center, Hirosawa 2-1, Wako, Saitama 351-0198, Japan}
\affiliation{Department of Physics, Kyoto University, Kita-shirakawa, Kyoto 606-8502, Japan}

\author{R.~Wang}%~(\CJKfamily{gbsn}{王})}
\affiliation{National Superconducting Cyclotron Laboratory, Michigan State University, East Lansing, Michigan 48824, USA}

\author{Dan~Cozma}
\affiliation{IFIN-HH, Reactorului 30, 077125 M\v{a}gurele-Bucharest, Romania}

\author{Rohit~Kumar}
\affiliation{National Superconducting Cyclotron Laboratory, Michigan State University, East Lansing, Michigan 48824, USA}

\author{Akira~Ono}
\affiliation{Department of Physics, Tohoku University, Sendai 980-8578, Japan}

\author{Ying-Xun~Zhang}
\affiliation{China Institute of Atomic Energy, Beijing 102413, China}

%\author{P.~Danielewicz}
%%\author{D.~Oliinychenko}
%\author{H.~Elfnerg}
%\author{N.~Ikeno}
%\author{C.M.~Ko}
%\author{J. Moh}
%\author{A.~Ono}
%\author{J.~Su}
%\author{Yong Jia Wang}
%\author{H.~Wolter}
%\author{Zhen Zhang}
%\author{Ying-Xun Zhang}
%\collaboration{TMEP collaboration}


%% file: output.bbl

%% file: apssamp.bbl
\begin{thebibliography}{0}%
\makeatletter
\providecommand \@ifxundefined [1]{%
 \@ifx{#1\undefined}
}%
\providecommand \@ifnum [1]{%
 \ifnum #1\expandafter \@firstoftwo
 \else \expandafter \@secondoftwo
 \fi
}%
\providecommand \@ifx [1]{%
 \ifx #1\expandafter \@firstoftwo
 \else \expandafter \@secondoftwo
 \fi
}%
\providecommand \natexlab [1]{#1}%
\providecommand \enquote  [1]{``#1''}%
\providecommand \bibnamefont  [1]{#1}%
\providecommand \bibfnamefont [1]{#1}%
\providecommand \citenamefont [1]{#1}%
\providecommand \href@noop [0]{\@secondoftwo}%
\providecommand \href [0]{\begingroup \@sanitize@url \@href}%
\providecommand \@href[1]{\@@startlink{#1}\@@href}%
\providecommand \@@href[1]{\endgroup#1\@@endlink}%
\providecommand \@sanitize@url [0]{\catcode `\\12\catcode `\$12\catcode
  `\&12\catcode `\#12\catcode `\^12\catcode `\_12\catcode `\%12\relax}%
\providecommand \@@startlink[1]{}%
\providecommand \@@endlink[0]{}%
\providecommand \url  [0]{\begingroup\@sanitize@url \@url }%
\providecommand \@url [1]{\endgroup\@href {#1}{\urlprefix }}%
\providecommand \urlprefix  [0]{URL }%
\providecommand \Eprint [0]{\href }%
\providecommand \doibase [0]{https://doi.org/}%
\providecommand \selectlanguage [0]{\@gobble}%
\providecommand \bibinfo  [0]{\@secondoftwo}%
\providecommand \bibfield  [0]{\@secondoftwo}%
\providecommand \translation [1]{[#1]}%
\providecommand \BibitemOpen [0]{}%
\providecommand \bibitemStop [0]{}%
\providecommand \bibitemNoStop [0]{.\EOS\space}%
\providecommand \EOS [0]{\spacefactor3000\relax}%
\providecommand \BibitemShut  [1]{\csname bibitem#1\endcsname}%
\let\auto@bib@innerbib\@empty
%</preamble>
\end{thebibliography}%


\begin{thebibliography}{32}%
\makeatletter
\providecommand \@ifxundefined [1]{%
 \@ifx{#1\undefined}
}%
\providecommand \@ifnum [1]{%
 \ifnum #1\expandafter \@firstoftwo
 \else \expandafter \@secondoftwo
 \fi
}%
\providecommand \@ifx [1]{%
 \ifx #1\expandafter \@firstoftwo
 \else \expandafter \@secondoftwo
 \fi
}%
\providecommand \natexlab [1]{#1}%
\providecommand \enquote  [1]{``#1''}%
\providecommand \bibnamefont  [1]{#1}%
\providecommand \bibfnamefont [1]{#1}%
\providecommand \citenamefont [1]{#1}%
\providecommand \href@noop [0]{\@secondoftwo}%
\providecommand \href [0]{\begingroup \@sanitize@url \@href}%
\providecommand \@href[1]{\@@startlink{#1}\@@href}%
\providecommand \@@href[1]{\endgroup#1\@@endlink}%
\providecommand \@sanitize@url [0]{\catcode `\\12\catcode `\$12\catcode
  `\&12\catcode `\#12\catcode `\^12\catcode `\_12\catcode `\%12\relax}%
\providecommand \@@startlink[1]{}%
\providecommand \@@endlink[0]{}%
\providecommand \url  [0]{\begingroup\@sanitize@url \@url }%
\providecommand \@url [1]{\endgroup\@href {#1}{\urlprefix }}%
\providecommand \urlprefix  [0]{URL }%
\providecommand \Eprint [0]{\href }%
\providecommand \doibase [0]{https://doi.org/}%
\providecommand \selectlanguage [0]{\@gobble}%
\providecommand \bibinfo  [0]{\@secondoftwo}%
\providecommand \bibfield  [0]{\@secondoftwo}%
\providecommand \translation [1]{[#1]}%
\providecommand \BibitemOpen [0]{}%
\providecommand \bibitemStop [0]{}%
\providecommand \bibitemNoStop [0]{.\EOS\space}%
\providecommand \EOS [0]{\spacefactor3000\relax}%
\providecommand \BibitemShut  [1]{\csname bibitem#1\endcsname}%
\let\auto@bib@innerbib\@empty
%</preamble>
\bibitem [{\citenamefont {{Deng}}(2012)}]{Deng12}%
  \BibitemOpen
  \bibfield  {author} {\bibinfo {author} {\bibfnamefont {L.}~\bibnamefont
  {{Deng}}},\ }\bibfield  {title} {\bibinfo {title} {The mnist database of
  handwritten digit images for machine learning research [best of the web]},\
  }\href {https://doi.org/10.1109/MSP.2012.2211477} {\bibfield  {journal}
  {\bibinfo  {journal} {IEEE Signal Processing Magazine}\ }\textbf {\bibinfo
  {volume} {29}},\ \bibinfo {pages} {141} (\bibinfo {year} {2012})}\BibitemShut
  {NoStop}%
\bibitem [{\citenamefont {Gader}\ \emph {et~al.}(1995)\citenamefont {Gader},
  \citenamefont {Whalen}, \citenamefont {Ganzberger},\ and\ \citenamefont
  {Hepp}}]{Gader95}%
  \BibitemOpen
  \bibfield  {author} {\bibinfo {author} {\bibfnamefont {P.}~\bibnamefont
  {Gader}}, \bibinfo {author} {\bibfnamefont {M.}~\bibnamefont {Whalen}},
  \bibinfo {author} {\bibfnamefont {M.}~\bibnamefont {Ganzberger}},\ and\
  \bibinfo {author} {\bibfnamefont {D.}~\bibnamefont {Hepp}},\ }\bibfield
  {title} {\bibinfo {title} {Handprinted word recognition on a nist data set},\
  }\href {https://doi.org/10.1007/BF01213636} {\bibfield  {journal} {\bibinfo
  {journal} {Machine Vision and Applications}\ }\textbf {\bibinfo {volume}
  {8}},\ \bibinfo {pages} {31} (\bibinfo {year} {1995})}\BibitemShut {NoStop}%
\bibitem [{\citenamefont {Estee}\ \emph {et~al.}(2021)\citenamefont {Estee},
  \citenamefont {Lynch}, \citenamefont {Tsang}, \citenamefont {Barney},
  \citenamefont {Jhang}, \citenamefont {Tsang}, \citenamefont {Wang},
  \citenamefont {Kaneko}, \citenamefont {Lee}, \citenamefont {Isobe},
  \citenamefont {Kurata-Nishimura}, \citenamefont {Murakami}, \citenamefont
  {Ahn}, \citenamefont {Atar}, \citenamefont {Aumann}, \citenamefont {Baba},
  \citenamefont {Boretzky}, \citenamefont {Brzychczyk}, \citenamefont
  {Cerizza}, \citenamefont {Chiga}, \citenamefont {Fukuda}, \citenamefont
  {Gasparic}, \citenamefont {Hong}, \citenamefont {Horvat}, \citenamefont
  {Ieki}, \citenamefont {Inabe}, \citenamefont {Kim}, \citenamefont
  {Kobayashi}, \citenamefont {Kondo}, \citenamefont {Lasko}, \citenamefont
  {Lee}, \citenamefont {Leifels}, \citenamefont {\L{}ukasik}, \citenamefont
  {Manfredi}, \citenamefont {McIntosh}, \citenamefont {Morfouace},
  \citenamefont {Nakamura}, \citenamefont {Nakatsuka}, \citenamefont
  {Nishimura}, \citenamefont {Otsu}, \citenamefont {Paw\l{}owski},
  \citenamefont {Pelczar}, \citenamefont {Rossi}, \citenamefont {Sakurai},
  \citenamefont {Santamaria}, \citenamefont {Sato}, \citenamefont {Scheit},
  \citenamefont {Shane}, \citenamefont {Shimizu}, \citenamefont {Simon},
  \citenamefont {Snoch}, \citenamefont {Sochocka}, \citenamefont {Sumikama},
  \citenamefont {Suzuki}, \citenamefont {Suzuki}, \citenamefont {Takeda},
  \citenamefont {Tangwancharoen}, \citenamefont {Toernqvist}, \citenamefont
  {Togano}, \citenamefont {Xiao}, \citenamefont {Yennello}, \citenamefont
  {Zhang},\ and\ \citenamefont {Cozma}}]{estee2021}%
  \BibitemOpen
  \bibfield  {author} {\bibinfo {author} {\bibfnamefont {J.}~\bibnamefont
  {Estee}}, \bibinfo {author} {\bibfnamefont {W.~G.}\ \bibnamefont {Lynch}},
  \bibinfo {author} {\bibfnamefont {C.~Y.}\ \bibnamefont {Tsang}}, \bibinfo
  {author} {\bibfnamefont {J.}~\bibnamefont {Barney}}, \bibinfo {author}
  {\bibfnamefont {G.}~\bibnamefont {Jhang}}, \bibinfo {author} {\bibfnamefont
  {M.~B.}\ \bibnamefont {Tsang}}, \bibinfo {author} {\bibfnamefont
  {R.}~\bibnamefont {Wang}}, \bibinfo {author} {\bibfnamefont {M.}~\bibnamefont
  {Kaneko}}, \bibinfo {author} {\bibfnamefont {J.~W.}\ \bibnamefont {Lee}},
  \bibinfo {author} {\bibfnamefont {T.}~\bibnamefont {Isobe}}, \bibinfo
  {author} {\bibfnamefont {M.}~\bibnamefont {Kurata-Nishimura}}, \bibinfo
  {author} {\bibfnamefont {T.}~\bibnamefont {Murakami}}, \bibinfo {author}
  {\bibfnamefont {D.~S.}\ \bibnamefont {Ahn}}, \bibinfo {author} {\bibfnamefont
  {L.}~\bibnamefont {Atar}}, \bibinfo {author} {\bibfnamefont {T.}~\bibnamefont
  {Aumann}}, \bibinfo {author} {\bibfnamefont {H.}~\bibnamefont {Baba}},
  \bibinfo {author} {\bibfnamefont {K.}~\bibnamefont {Boretzky}}, \bibinfo
  {author} {\bibfnamefont {J.}~\bibnamefont {Brzychczyk}}, \bibinfo {author}
  {\bibfnamefont {G.}~\bibnamefont {Cerizza}}, \bibinfo {author} {\bibfnamefont
  {N.}~\bibnamefont {Chiga}}, \bibinfo {author} {\bibfnamefont
  {N.}~\bibnamefont {Fukuda}}, \bibinfo {author} {\bibfnamefont
  {I.}~\bibnamefont {Gasparic}}, \bibinfo {author} {\bibfnamefont
  {B.}~\bibnamefont {Hong}}, \bibinfo {author} {\bibfnamefont {A.}~\bibnamefont
  {Horvat}}, \bibinfo {author} {\bibfnamefont {K.}~\bibnamefont {Ieki}},
  \bibinfo {author} {\bibfnamefont {N.}~\bibnamefont {Inabe}}, \bibinfo
  {author} {\bibfnamefont {Y.~J.}\ \bibnamefont {Kim}}, \bibinfo {author}
  {\bibfnamefont {T.}~\bibnamefont {Kobayashi}}, \bibinfo {author}
  {\bibfnamefont {Y.}~\bibnamefont {Kondo}}, \bibinfo {author} {\bibfnamefont
  {P.}~\bibnamefont {Lasko}}, \bibinfo {author} {\bibfnamefont {H.~S.}\
  \bibnamefont {Lee}}, \bibinfo {author} {\bibfnamefont {Y.}~\bibnamefont
  {Leifels}}, \bibinfo {author} {\bibfnamefont {J.}~\bibnamefont {\L{}ukasik}},
  \bibinfo {author} {\bibfnamefont {J.}~\bibnamefont {Manfredi}}, \bibinfo
  {author} {\bibfnamefont {A.~B.}\ \bibnamefont {McIntosh}}, \bibinfo {author}
  {\bibfnamefont {P.}~\bibnamefont {Morfouace}}, \bibinfo {author}
  {\bibfnamefont {T.}~\bibnamefont {Nakamura}}, \bibinfo {author}
  {\bibfnamefont {N.}~\bibnamefont {Nakatsuka}}, \bibinfo {author}
  {\bibfnamefont {S.}~\bibnamefont {Nishimura}}, \bibinfo {author}
  {\bibfnamefont {H.}~\bibnamefont {Otsu}}, \bibinfo {author} {\bibfnamefont
  {P.}~\bibnamefont {Paw\l{}owski}}, \bibinfo {author} {\bibfnamefont
  {K.}~\bibnamefont {Pelczar}}, \bibinfo {author} {\bibfnamefont
  {D.}~\bibnamefont {Rossi}}, \bibinfo {author} {\bibfnamefont
  {H.}~\bibnamefont {Sakurai}}, \bibinfo {author} {\bibfnamefont
  {C.}~\bibnamefont {Santamaria}}, \bibinfo {author} {\bibfnamefont
  {H.}~\bibnamefont {Sato}}, \bibinfo {author} {\bibfnamefont {H.}~\bibnamefont
  {Scheit}}, \bibinfo {author} {\bibfnamefont {R.}~\bibnamefont {Shane}},
  \bibinfo {author} {\bibfnamefont {Y.}~\bibnamefont {Shimizu}}, \bibinfo
  {author} {\bibfnamefont {H.}~\bibnamefont {Simon}}, \bibinfo {author}
  {\bibfnamefont {A.}~\bibnamefont {Snoch}}, \bibinfo {author} {\bibfnamefont
  {A.}~\bibnamefont {Sochocka}}, \bibinfo {author} {\bibfnamefont
  {T.}~\bibnamefont {Sumikama}}, \bibinfo {author} {\bibfnamefont
  {H.}~\bibnamefont {Suzuki}}, \bibinfo {author} {\bibfnamefont
  {D.}~\bibnamefont {Suzuki}}, \bibinfo {author} {\bibfnamefont
  {H.}~\bibnamefont {Takeda}}, \bibinfo {author} {\bibfnamefont
  {S.}~\bibnamefont {Tangwancharoen}}, \bibinfo {author} {\bibfnamefont
  {H.}~\bibnamefont {Toernqvist}}, \bibinfo {author} {\bibfnamefont
  {Y.}~\bibnamefont {Togano}}, \bibinfo {author} {\bibfnamefont {Z.~G.}\
  \bibnamefont {Xiao}}, \bibinfo {author} {\bibfnamefont {S.~J.}\ \bibnamefont
  {Yennello}}, \bibinfo {author} {\bibfnamefont {Y.}~\bibnamefont {Zhang}},\
  and\ \bibinfo {author} {\bibfnamefont {M.~D.}\ \bibnamefont {Cozma}}
  (\bibinfo {collaboration} {$\mathrm{S}\ensuremath{\pi}\mathrm{RIT}$
  Collaboration}),\ }\bibfield  {title} {\bibinfo {title} {Probing the symmetry
  energy with the spectral pion ratio},\ }\href
  {https://doi.org/10.1103/PhysRevLett.126.162701} {\bibfield  {journal}
  {\bibinfo  {journal} {Phys. Rev. Lett.}\ }\textbf {\bibinfo {volume} {126}},\
  \bibinfo {pages} {162701} (\bibinfo {year} {2021})}\BibitemShut {NoStop}%
\bibitem [{\citenamefont {Jhang}\ \emph {et~al.}(2021)\citenamefont {Jhang},
  \citenamefont {Estee}, \citenamefont {Barney}, \citenamefont {Cerizza},
  \citenamefont {Kaneko}, \citenamefont {Lee}, \citenamefont {Lynch},
  \citenamefont {Isobe}, \citenamefont {Kurata-Nishimura}, \citenamefont
  {Murakami}, \citenamefont {Tsang}, \citenamefont {Tsang}, \citenamefont
  {Wang}, \citenamefont {Ahn}, \citenamefont {Atar}, \citenamefont {Aumann},
  \citenamefont {Baba}, \citenamefont {Boretzky}, \citenamefont {Brzychczyk},
  \citenamefont {Chiga}, \citenamefont {Fukuda}, \citenamefont {Gasparic},
  \citenamefont {Hong}, \citenamefont {Horvat}, \citenamefont {Ieki},
  \citenamefont {Inabe}, \citenamefont {Kim}, \citenamefont {Kobayashi},
  \citenamefont {Kondo}, \citenamefont {Lasko}, \citenamefont {Lee},
  \citenamefont {Leifels}, \citenamefont {Łukasik}, \citenamefont {Manfredi},
  \citenamefont {McIntosh}, \citenamefont {Morfouace}, \citenamefont
  {Nakamura}, \citenamefont {Nakatsuka}, \citenamefont {Nishimura},
  \citenamefont {Olsen}, \citenamefont {Otsu}, \citenamefont {Pawłowski},
  \citenamefont {Pelczar}, \citenamefont {Rossi}, \citenamefont {Sakurai},
  \citenamefont {Santamaria}, \citenamefont {Sato}, \citenamefont {Scheit},
  \citenamefont {Shane}, \citenamefont {Shimizu}, \citenamefont {Simon},
  \citenamefont {Snoch}, \citenamefont {Sochocka}, \citenamefont {Sosin},
  \citenamefont {Sumikama}, \citenamefont {Suzuki}, \citenamefont {Suzuki},
  \citenamefont {Takeda}, \citenamefont {Tangwancharoen}, \citenamefont
  {Toernqvist}, \citenamefont {Togano}, \citenamefont {Xiao}, \citenamefont
  {Yennello}, \citenamefont {Yurkon}, \citenamefont {Zhang}, \citenamefont
  {Colonna}, \citenamefont {Cozma}, \citenamefont {Danielewicz}, \citenamefont
  {Elfner}, \citenamefont {Ikeno}, \citenamefont {Ko}, \citenamefont {Mohs},
  \citenamefont {Oliinychenko}, \citenamefont {Ono}, \citenamefont {Su},
  \citenamefont {Wang}, \citenamefont {Wolter}, \citenamefont {Xu},
  \citenamefont {Zhang},\ and\ \citenamefont {Zhang}}]{Jhang21}%
  \BibitemOpen
  \bibfield  {author} {\bibinfo {author} {\bibfnamefont {G.}~\bibnamefont
  {Jhang}}, \bibinfo {author} {\bibfnamefont {J.}~\bibnamefont {Estee}},
  \bibinfo {author} {\bibfnamefont {J.}~\bibnamefont {Barney}}, \bibinfo
  {author} {\bibfnamefont {G.}~\bibnamefont {Cerizza}}, \bibinfo {author}
  {\bibfnamefont {M.}~\bibnamefont {Kaneko}}, \bibinfo {author} {\bibfnamefont
  {J.}~\bibnamefont {Lee}}, \bibinfo {author} {\bibfnamefont {W.}~\bibnamefont
  {Lynch}}, \bibinfo {author} {\bibfnamefont {T.}~\bibnamefont {Isobe}},
  \bibinfo {author} {\bibfnamefont {M.}~\bibnamefont {Kurata-Nishimura}},
  \bibinfo {author} {\bibfnamefont {T.}~\bibnamefont {Murakami}}, \bibinfo
  {author} {\bibfnamefont {C.}~\bibnamefont {Tsang}}, \bibinfo {author}
  {\bibfnamefont {M.}~\bibnamefont {Tsang}}, \bibinfo {author} {\bibfnamefont
  {R.}~\bibnamefont {Wang}}, \bibinfo {author} {\bibfnamefont {D.}~\bibnamefont
  {Ahn}}, \bibinfo {author} {\bibfnamefont {L.}~\bibnamefont {Atar}}, \bibinfo
  {author} {\bibfnamefont {T.}~\bibnamefont {Aumann}}, \bibinfo {author}
  {\bibfnamefont {H.}~\bibnamefont {Baba}}, \bibinfo {author} {\bibfnamefont
  {K.}~\bibnamefont {Boretzky}}, \bibinfo {author} {\bibfnamefont
  {J.}~\bibnamefont {Brzychczyk}}, \bibinfo {author} {\bibfnamefont
  {N.}~\bibnamefont {Chiga}}, \bibinfo {author} {\bibfnamefont
  {N.}~\bibnamefont {Fukuda}}, \bibinfo {author} {\bibfnamefont
  {I.}~\bibnamefont {Gasparic}}, \bibinfo {author} {\bibfnamefont
  {B.}~\bibnamefont {Hong}}, \bibinfo {author} {\bibfnamefont {A.}~\bibnamefont
  {Horvat}}, \bibinfo {author} {\bibfnamefont {K.}~\bibnamefont {Ieki}},
  \bibinfo {author} {\bibfnamefont {N.}~\bibnamefont {Inabe}}, \bibinfo
  {author} {\bibfnamefont {Y.}~\bibnamefont {Kim}}, \bibinfo {author}
  {\bibfnamefont {T.}~\bibnamefont {Kobayashi}}, \bibinfo {author}
  {\bibfnamefont {Y.}~\bibnamefont {Kondo}}, \bibinfo {author} {\bibfnamefont
  {P.}~\bibnamefont {Lasko}}, \bibinfo {author} {\bibfnamefont
  {H.}~\bibnamefont {Lee}}, \bibinfo {author} {\bibfnamefont {Y.}~\bibnamefont
  {Leifels}}, \bibinfo {author} {\bibfnamefont {J.}~\bibnamefont {Łukasik}},
  \bibinfo {author} {\bibfnamefont {J.}~\bibnamefont {Manfredi}}, \bibinfo
  {author} {\bibfnamefont {A.}~\bibnamefont {McIntosh}}, \bibinfo {author}
  {\bibfnamefont {P.}~\bibnamefont {Morfouace}}, \bibinfo {author}
  {\bibfnamefont {T.}~\bibnamefont {Nakamura}}, \bibinfo {author}
  {\bibfnamefont {N.}~\bibnamefont {Nakatsuka}}, \bibinfo {author}
  {\bibfnamefont {S.}~\bibnamefont {Nishimura}}, \bibinfo {author}
  {\bibfnamefont {R.}~\bibnamefont {Olsen}}, \bibinfo {author} {\bibfnamefont
  {H.}~\bibnamefont {Otsu}}, \bibinfo {author} {\bibfnamefont {P.}~\bibnamefont
  {Pawłowski}}, \bibinfo {author} {\bibfnamefont {K.}~\bibnamefont {Pelczar}},
  \bibinfo {author} {\bibfnamefont {D.}~\bibnamefont {Rossi}}, \bibinfo
  {author} {\bibfnamefont {H.}~\bibnamefont {Sakurai}}, \bibinfo {author}
  {\bibfnamefont {C.}~\bibnamefont {Santamaria}}, \bibinfo {author}
  {\bibfnamefont {H.}~\bibnamefont {Sato}}, \bibinfo {author} {\bibfnamefont
  {H.}~\bibnamefont {Scheit}}, \bibinfo {author} {\bibfnamefont
  {R.}~\bibnamefont {Shane}}, \bibinfo {author} {\bibfnamefont
  {Y.}~\bibnamefont {Shimizu}}, \bibinfo {author} {\bibfnamefont
  {H.}~\bibnamefont {Simon}}, \bibinfo {author} {\bibfnamefont
  {A.}~\bibnamefont {Snoch}}, \bibinfo {author} {\bibfnamefont
  {A.}~\bibnamefont {Sochocka}}, \bibinfo {author} {\bibfnamefont
  {Z.}~\bibnamefont {Sosin}}, \bibinfo {author} {\bibfnamefont
  {T.}~\bibnamefont {Sumikama}}, \bibinfo {author} {\bibfnamefont
  {H.}~\bibnamefont {Suzuki}}, \bibinfo {author} {\bibfnamefont
  {D.}~\bibnamefont {Suzuki}}, \bibinfo {author} {\bibfnamefont
  {H.}~\bibnamefont {Takeda}}, \bibinfo {author} {\bibfnamefont
  {S.}~\bibnamefont {Tangwancharoen}}, \bibinfo {author} {\bibfnamefont
  {H.}~\bibnamefont {Toernqvist}}, \bibinfo {author} {\bibfnamefont
  {Y.}~\bibnamefont {Togano}}, \bibinfo {author} {\bibfnamefont
  {Z.}~\bibnamefont {Xiao}}, \bibinfo {author} {\bibfnamefont {S.}~\bibnamefont
  {Yennello}}, \bibinfo {author} {\bibfnamefont {J.}~\bibnamefont {Yurkon}},
  \bibinfo {author} {\bibfnamefont {Y.}~\bibnamefont {Zhang}}, \bibinfo
  {author} {\bibfnamefont {M.}~\bibnamefont {Colonna}}, \bibinfo {author}
  {\bibfnamefont {D.}~\bibnamefont {Cozma}}, \bibinfo {author} {\bibfnamefont
  {P.}~\bibnamefont {Danielewicz}}, \bibinfo {author} {\bibfnamefont
  {H.}~\bibnamefont {Elfner}}, \bibinfo {author} {\bibfnamefont
  {N.}~\bibnamefont {Ikeno}}, \bibinfo {author} {\bibfnamefont {C.~M.}\
  \bibnamefont {Ko}}, \bibinfo {author} {\bibfnamefont {J.}~\bibnamefont
  {Mohs}}, \bibinfo {author} {\bibfnamefont {D.}~\bibnamefont {Oliinychenko}},
  \bibinfo {author} {\bibfnamefont {A.}~\bibnamefont {Ono}}, \bibinfo {author}
  {\bibfnamefont {J.}~\bibnamefont {Su}}, \bibinfo {author} {\bibfnamefont
  {Y.~J.}\ \bibnamefont {Wang}}, \bibinfo {author} {\bibfnamefont
  {H.}~\bibnamefont {Wolter}}, \bibinfo {author} {\bibfnamefont
  {J.}~\bibnamefont {Xu}}, \bibinfo {author} {\bibfnamefont {Y.-X.}\
  \bibnamefont {Zhang}},\ and\ \bibinfo {author} {\bibfnamefont
  {Z.}~\bibnamefont {Zhang}},\ }\bibfield  {title} {\bibinfo {title} {{Symmetry
  energy investigation with pion production from Sn+Sn systems}},\ }\href
  {https://doi.org/https://doi.org/10.1016/j.physletb.2020.136016} {\bibfield
  {journal} {\bibinfo  {journal} {Phys. Lett. B}\ }\textbf {\bibinfo {volume}
  {813}},\ \bibinfo {pages} {136016} (\bibinfo {year} {2021})}\BibitemShut
  {NoStop}%
\bibitem [{\citenamefont {Bass}\ \emph {et~al.}(1994)\citenamefont {Bass},
  \citenamefont {Bischoff}, \citenamefont {Hartnack}, \citenamefont {Maruhn},
  \citenamefont {Reinhardt}, \citenamefont {Stocker},\ and\ \citenamefont
  {Greiner}}]{Bass94}%
  \BibitemOpen
  \bibfield  {author} {\bibinfo {author} {\bibfnamefont {S.~A.}\ \bibnamefont
  {Bass}}, \bibinfo {author} {\bibfnamefont {A.}~\bibnamefont {Bischoff}},
  \bibinfo {author} {\bibfnamefont {C.}~\bibnamefont {Hartnack}}, \bibinfo
  {author} {\bibfnamefont {J.~A.}\ \bibnamefont {Maruhn}}, \bibinfo {author}
  {\bibfnamefont {J.}~\bibnamefont {Reinhardt}}, \bibinfo {author}
  {\bibfnamefont {H.}~\bibnamefont {Stocker}},\ and\ \bibinfo {author}
  {\bibfnamefont {W.}~\bibnamefont {Greiner}},\ }\bibfield  {title} {\bibinfo
  {title} {Neural networks for impact parameter determination},\ }\href
  {https://doi.org/10.1088/0954-3899/20/1/004} {\bibfield  {journal} {\bibinfo
  {journal} {J. of Phys. G: Nucl. and Part. Phys.}\ }\textbf {\bibinfo {volume}
  {20}},\ \bibinfo {pages} {L21} (\bibinfo {year} {1994})}\BibitemShut
  {NoStop}%
\bibitem [{\citenamefont {Bass}\ \emph {et~al.}(1996)\citenamefont {Bass},
  \citenamefont {Bischoff}, \citenamefont {Maruhn}, \citenamefont {St\"ocker},\
  and\ \citenamefont {Greiner}}]{Bass96}%
  \BibitemOpen
  \bibfield  {author} {\bibinfo {author} {\bibfnamefont {S.~A.}\ \bibnamefont
  {Bass}}, \bibinfo {author} {\bibfnamefont {A.}~\bibnamefont {Bischoff}},
  \bibinfo {author} {\bibfnamefont {J.~A.}\ \bibnamefont {Maruhn}}, \bibinfo
  {author} {\bibfnamefont {H.}~\bibnamefont {St\"ocker}},\ and\ \bibinfo
  {author} {\bibfnamefont {W.}~\bibnamefont {Greiner}},\ }\bibfield  {title}
  {\bibinfo {title} {{Neural networks for impact parameter determination}},\
  }\href {https://doi.org/10.1103/PhysRevC.53.2358} {\bibfield  {journal}
  {\bibinfo  {journal} {Phys. Rev. C}\ }\textbf {\bibinfo {volume} {53}},\
  \bibinfo {pages} {2358} (\bibinfo {year} {1996})}\BibitemShut {NoStop}%
\bibitem [{\citenamefont {Sanctis}\ \emph {et~al.}(2008)\citenamefont
  {Sanctis}, \citenamefont {Masotti}, \citenamefont {Bruno}, \citenamefont
  {D{\textquotesingle}Agostino}, \citenamefont {Geraci}, \citenamefont
  {Vannini},\ and\ \citenamefont {Bonasera}}]{De_Sanctis_2008}%
  \BibitemOpen
  \bibfield  {author} {\bibinfo {author} {\bibfnamefont {J.~D.}\ \bibnamefont
  {Sanctis}}, \bibinfo {author} {\bibfnamefont {M.}~\bibnamefont {Masotti}},
  \bibinfo {author} {\bibfnamefont {M.}~\bibnamefont {Bruno}}, \bibinfo
  {author} {\bibfnamefont {M.}~\bibnamefont {D{\textquotesingle}Agostino}},
  \bibinfo {author} {\bibfnamefont {E.}~\bibnamefont {Geraci}}, \bibinfo
  {author} {\bibfnamefont {G.}~\bibnamefont {Vannini}},\ and\ \bibinfo {author}
  {\bibfnamefont {A.}~\bibnamefont {Bonasera}},\ }\bibfield  {title} {\bibinfo
  {title} {Classification of the impact parameter in
  nucleus{\textendash}nucleus collisions by a support vector machine method},\
  }\href {https://doi.org/10.1088/0954-3899/36/1/015101} {\bibfield  {journal}
  {\bibinfo  {journal} {J. of Phys. G: Nucl. and Part. Phys.}\ }\textbf
  {\bibinfo {volume} {36}},\ \bibinfo {pages} {015101} (\bibinfo {year}
  {2008})}\BibitemShut {NoStop}%
\bibitem [{\citenamefont {David}\ \emph {et~al.}(1995)\citenamefont {David},
  \citenamefont {Freslier},\ and\ \citenamefont {Aichelin}}]{Christophe1995}%
  \BibitemOpen
  \bibfield  {author} {\bibinfo {author} {\bibfnamefont {C.}~\bibnamefont
  {David}}, \bibinfo {author} {\bibfnamefont {M.}~\bibnamefont {Freslier}},\
  and\ \bibinfo {author} {\bibfnamefont {J.}~\bibnamefont {Aichelin}},\
  }\bibfield  {title} {\bibinfo {title} {Impact parameter determination for
  heavy-ion collisions by use of a neural network},\ }\href
  {https://doi.org/10.1103/PhysRevC.51.1453} {\bibfield  {journal} {\bibinfo
  {journal} {Phys. Rev. C}\ }\textbf {\bibinfo {volume} {51}},\ \bibinfo
  {pages} {1453} (\bibinfo {year} {1995})}\BibitemShut {NoStop}%
\bibitem [{\citenamefont {{Omana Kuttan}}\ \emph {et~al.}(2020)\citenamefont
  {{Omana Kuttan}}, \citenamefont {Steinheimer}, \citenamefont {Zhou},
  \citenamefont {Redelbach},\ and\ \citenamefont {Stoecker}}]{Oma19}%
  \BibitemOpen
  \bibfield  {author} {\bibinfo {author} {\bibfnamefont {M.}~\bibnamefont
  {{Omana Kuttan}}}, \bibinfo {author} {\bibfnamefont {J.}~\bibnamefont
  {Steinheimer}}, \bibinfo {author} {\bibfnamefont {K.}~\bibnamefont {Zhou}},
  \bibinfo {author} {\bibfnamefont {A.}~\bibnamefont {Redelbach}},\ and\
  \bibinfo {author} {\bibfnamefont {H.}~\bibnamefont {Stoecker}},\ }\bibfield
  {title} {\bibinfo {title} {A fast centrality-meter for heavy-ion collisions
  at the cbm experiment},\ }\href
  {https://doi.org/https://doi.org/10.1016/j.physletb.2020.135872} {\bibfield
  {journal} {\bibinfo  {journal} {Phys. Lett. B}\ }\textbf {\bibinfo {volume}
  {811}},\ \bibinfo {pages} {135872} (\bibinfo {year} {2020})}\BibitemShut
  {NoStop}%
\bibitem [{\citenamefont {Wang}\ \emph {et~al.}(2020)\citenamefont {Wang},
  \citenamefont {Li}, \citenamefont {Leifels},\ and\ \citenamefont {{Le
  Fèvre}}}]{WANG2020}%
  \BibitemOpen
  \bibfield  {author} {\bibinfo {author} {\bibfnamefont {Y.}~\bibnamefont
  {Wang}}, \bibinfo {author} {\bibfnamefont {Q.}~\bibnamefont {Li}}, \bibinfo
  {author} {\bibfnamefont {Y.}~\bibnamefont {Leifels}},\ and\ \bibinfo {author}
  {\bibfnamefont {A.}~\bibnamefont {{Le Fèvre}}},\ }\bibfield  {title}
  {\bibinfo {title} {{Study of the nuclear symmetry energy from the
  rapidity-dependent elliptic flow in heavy-ion collisions around 1 GeV/nucleon
  regime}},\ }\href
  {https://doi.org/https://doi.org/10.1016/j.physletb.2020.135249} {\bibfield
  {journal} {\bibinfo  {journal} {Phys. Lett. B}\ }\textbf {\bibinfo {volume}
  {802}},\ \bibinfo {pages} {135249} (\bibinfo {year} {2020})}\BibitemShut
  {NoStop}%
\bibitem [{\citenamefont {Li}\ \emph {et~al.}(2021)\citenamefont {Li},
  \citenamefont {Wang}, \citenamefont {Gao}, \citenamefont {Li}, \citenamefont
  {Lv}, \citenamefont {Li}, \citenamefont {Tsang},\ and\ \citenamefont
  {Tsang}}]{li2021application}%
  \BibitemOpen
  \bibfield  {author} {\bibinfo {author} {\bibfnamefont {F.}~\bibnamefont
  {Li}}, \bibinfo {author} {\bibfnamefont {Y.}~\bibnamefont {Wang}}, \bibinfo
  {author} {\bibfnamefont {Z.}~\bibnamefont {Gao}}, \bibinfo {author}
  {\bibfnamefont {P.}~\bibnamefont {Li}}, \bibinfo {author} {\bibfnamefont
  {H.}~\bibnamefont {Lv}}, \bibinfo {author} {\bibfnamefont {Q.}~\bibnamefont
  {Li}}, \bibinfo {author} {\bibfnamefont {C.~Y.}\ \bibnamefont {Tsang}},\ and\
  \bibinfo {author} {\bibfnamefont {M.~B.}\ \bibnamefont {Tsang}},\ }\href@noop
  {} {\bibinfo {title} {{Application of machine learning in the determination
  of impact parameter in the $^{132}$Sn+$^{124}$Sn system}}} (\bibinfo {year}
  {2021}),\ \Eprint {https://arxiv.org/abs/2105.08912} {arXiv:2105.08912
  [nucl-th]} \BibitemShut {NoStop}%
\bibitem [{\citenamefont {Estee}\ \emph {et~al.}(2019)\citenamefont {Estee},
  \citenamefont {Lynch}, \citenamefont {Barney}, \citenamefont {Cerizza},
  \citenamefont {Jhang}, \citenamefont {Lee}, \citenamefont {Wang},
  \citenamefont {Isobe}, \citenamefont {Kaneko}, \citenamefont
  {Kurata-Nishimura}, \citenamefont {Murakami}, \citenamefont {Shane},
  \citenamefont {Tangwancharoen}, \citenamefont {Tsang}, \citenamefont {Tsang},
  \citenamefont {Hong}, \citenamefont {Lasko}, \citenamefont {Łukasik},
  \citenamefont {McIntosh}, \citenamefont {Pawłowski}, \citenamefont
  {Pelczar}, \citenamefont {Sakurai}, \citenamefont {Santamaria}, \citenamefont
  {Suzuki}, \citenamefont {Yennello},\ and\ \citenamefont {Zhang}}]{Estee19}%
  \BibitemOpen
  \bibfield  {author} {\bibinfo {author} {\bibfnamefont {J.}~\bibnamefont
  {Estee}}, \bibinfo {author} {\bibfnamefont {W.}~\bibnamefont {Lynch}},
  \bibinfo {author} {\bibfnamefont {J.}~\bibnamefont {Barney}}, \bibinfo
  {author} {\bibfnamefont {G.}~\bibnamefont {Cerizza}}, \bibinfo {author}
  {\bibfnamefont {G.}~\bibnamefont {Jhang}}, \bibinfo {author} {\bibfnamefont
  {J.}~\bibnamefont {Lee}}, \bibinfo {author} {\bibfnamefont {R.}~\bibnamefont
  {Wang}}, \bibinfo {author} {\bibfnamefont {T.}~\bibnamefont {Isobe}},
  \bibinfo {author} {\bibfnamefont {M.}~\bibnamefont {Kaneko}}, \bibinfo
  {author} {\bibfnamefont {M.}~\bibnamefont {Kurata-Nishimura}}, \bibinfo
  {author} {\bibfnamefont {T.}~\bibnamefont {Murakami}}, \bibinfo {author}
  {\bibfnamefont {R.}~\bibnamefont {Shane}}, \bibinfo {author} {\bibfnamefont
  {S.}~\bibnamefont {Tangwancharoen}}, \bibinfo {author} {\bibfnamefont
  {C.}~\bibnamefont {Tsang}}, \bibinfo {author} {\bibfnamefont
  {M.}~\bibnamefont {Tsang}}, \bibinfo {author} {\bibfnamefont
  {B.}~\bibnamefont {Hong}}, \bibinfo {author} {\bibfnamefont {P.}~\bibnamefont
  {Lasko}}, \bibinfo {author} {\bibfnamefont {J.}~\bibnamefont {Łukasik}},
  \bibinfo {author} {\bibfnamefont {A.}~\bibnamefont {McIntosh}}, \bibinfo
  {author} {\bibfnamefont {P.}~\bibnamefont {Pawłowski}}, \bibinfo {author}
  {\bibfnamefont {K.}~\bibnamefont {Pelczar}}, \bibinfo {author} {\bibfnamefont
  {H.}~\bibnamefont {Sakurai}}, \bibinfo {author} {\bibfnamefont
  {C.}~\bibnamefont {Santamaria}}, \bibinfo {author} {\bibfnamefont
  {D.}~\bibnamefont {Suzuki}}, \bibinfo {author} {\bibfnamefont
  {S.}~\bibnamefont {Yennello}},\ and\ \bibinfo {author} {\bibfnamefont
  {Y.}~\bibnamefont {Zhang}},\ }\bibfield  {title} {\bibinfo {title} {Extending
  the dynamic range of electronics in a time projection chamber},\ }\href
  {https://doi.org/https://doi.org/10.1016/j.nima.2019.162509} {\bibfield
  {journal} {\bibinfo  {journal} {Nucl. Instru. Methods in Phys. Res.}\
  }\textbf {\bibinfo {volume} {944}},\ \bibinfo {pages} {162509} (\bibinfo
  {year} {2019})}\BibitemShut {NoStop}%
\bibitem [{\citenamefont {{Barney}}\ \emph {et~al.}(2020)\citenamefont
  {{Barney}}, \citenamefont {{Estee}}, \citenamefont {{Lynch}}, \citenamefont
  {{Isobe}}, \citenamefont {{Jhang}}, \citenamefont {{Kurata-Nishimura}},
  \citenamefont {{McIntosh}}, \citenamefont {{Murakami}}, \citenamefont
  {{Shane}}, \citenamefont {{Tangwancharoen}}, \citenamefont {{Tsang}},
  \citenamefont {{Cerizza}}, \citenamefont {{Kaneko}}, \citenamefont {{Lee}},
  \citenamefont {{Tsang}}, \citenamefont {{Wang}}, \citenamefont {{Anderson}},
  \citenamefont {{Baba}}, \citenamefont {{Chajecki}}, \citenamefont
  {{Famiano}}, \citenamefont {{Hodges-Showalter}}, \citenamefont {{Hong}},
  \citenamefont {{Kobayashi}}, \citenamefont {{Lasko}}, \citenamefont
  {{{\L}ukasik}}, \citenamefont {{Nakatsuka}}, \citenamefont {{Olsen}},
  \citenamefont {{Otsu}}, \citenamefont {{Paw{\l}owski}}, \citenamefont
  {{Pelczar}}, \citenamefont {{Powell}}, \citenamefont {{Sakurai}},
  \citenamefont {{Santamaria}}, \citenamefont {{Setiawan}}, \citenamefont
  {{Taketani}}, \citenamefont {{Winkelbauer}}, \citenamefont {{Xiao}},
  \citenamefont {{Yennello}}, \citenamefont {{Yurkon}},\ and\ \citenamefont
  {{Zhang}}}]{Barney20}%
  \BibitemOpen
  \bibfield  {author} {\bibinfo {author} {\bibfnamefont {J.}~\bibnamefont
  {{Barney}}}, \bibinfo {author} {\bibfnamefont {J.}~\bibnamefont {{Estee}}},
  \bibinfo {author} {\bibfnamefont {W.~G.}\ \bibnamefont {{Lynch}}}, \bibinfo
  {author} {\bibfnamefont {T.}~\bibnamefont {{Isobe}}}, \bibinfo {author}
  {\bibfnamefont {G.}~\bibnamefont {{Jhang}}}, \bibinfo {author} {\bibfnamefont
  {M.}~\bibnamefont {{Kurata-Nishimura}}}, \bibinfo {author} {\bibfnamefont
  {A.~B.}\ \bibnamefont {{McIntosh}}}, \bibinfo {author} {\bibfnamefont
  {T.}~\bibnamefont {{Murakami}}}, \bibinfo {author} {\bibfnamefont
  {R.}~\bibnamefont {{Shane}}}, \bibinfo {author} {\bibfnamefont
  {S.}~\bibnamefont {{Tangwancharoen}}}, \bibinfo {author} {\bibfnamefont
  {M.~B.}\ \bibnamefont {{Tsang}}}, \bibinfo {author} {\bibfnamefont
  {G.}~\bibnamefont {{Cerizza}}}, \bibinfo {author} {\bibfnamefont
  {M.}~\bibnamefont {{Kaneko}}}, \bibinfo {author} {\bibfnamefont {J.~W.}\
  \bibnamefont {{Lee}}}, \bibinfo {author} {\bibfnamefont {C.~Y.}\ \bibnamefont
  {{Tsang}}}, \bibinfo {author} {\bibfnamefont {R.}~\bibnamefont {{Wang}}},
  \bibinfo {author} {\bibfnamefont {C.}~\bibnamefont {{Anderson}}}, \bibinfo
  {author} {\bibfnamefont {H.}~\bibnamefont {{Baba}}}, \bibinfo {author}
  {\bibfnamefont {Z.}~\bibnamefont {{Chajecki}}}, \bibinfo {author}
  {\bibfnamefont {M.}~\bibnamefont {{Famiano}}}, \bibinfo {author}
  {\bibfnamefont {R.}~\bibnamefont {{Hodges-Showalter}}}, \bibinfo {author}
  {\bibfnamefont {B.}~\bibnamefont {{Hong}}}, \bibinfo {author} {\bibfnamefont
  {T.}~\bibnamefont {{Kobayashi}}}, \bibinfo {author} {\bibfnamefont
  {P.}~\bibnamefont {{Lasko}}}, \bibinfo {author} {\bibfnamefont
  {J.}~\bibnamefont {{{\L}ukasik}}}, \bibinfo {author} {\bibfnamefont
  {N.}~\bibnamefont {{Nakatsuka}}}, \bibinfo {author} {\bibfnamefont
  {R.}~\bibnamefont {{Olsen}}}, \bibinfo {author} {\bibfnamefont
  {H.}~\bibnamefont {{Otsu}}}, \bibinfo {author} {\bibfnamefont
  {P.}~\bibnamefont {{Paw{\l}owski}}}, \bibinfo {author} {\bibfnamefont
  {K.}~\bibnamefont {{Pelczar}}}, \bibinfo {author} {\bibfnamefont
  {W.}~\bibnamefont {{Powell}}}, \bibinfo {author} {\bibfnamefont
  {H.}~\bibnamefont {{Sakurai}}}, \bibinfo {author} {\bibfnamefont
  {C.}~\bibnamefont {{Santamaria}}}, \bibinfo {author} {\bibfnamefont
  {H.}~\bibnamefont {{Setiawan}}}, \bibinfo {author} {\bibfnamefont
  {A.}~\bibnamefont {{Taketani}}}, \bibinfo {author} {\bibfnamefont {J.~R.}\
  \bibnamefont {{Winkelbauer}}}, \bibinfo {author} {\bibfnamefont
  {Z.}~\bibnamefont {{Xiao}}}, \bibinfo {author} {\bibfnamefont {S.~J.}\
  \bibnamefont {{Yennello}}}, \bibinfo {author} {\bibfnamefont
  {J.}~\bibnamefont {{Yurkon}}},\ and\ \bibinfo {author} {\bibfnamefont
  {Y.}~\bibnamefont {{Zhang}}},\ }\bibfield  {title} {\bibinfo {title} {{The
  S$\pi$RIT Time Projection Chamber}},\ }\href@noop {} {\bibfield  {journal}
  {\bibinfo  {journal} {arXiv e-prints}\ ,\ \bibinfo {eid} {arXiv:2005.10806}}
  (\bibinfo {year} {2020})},\ \Eprint {https://arxiv.org/abs/2005.10806}
  {arXiv:2005.10806 [physics.ins-det]} \BibitemShut {NoStop}%
\bibitem [{\citenamefont {Kobayashi}\ \emph {et~al.}(2013)\citenamefont
  {Kobayashi}, \citenamefont {Chiga}, \citenamefont {Isobe}, \citenamefont
  {Kondo}, \citenamefont {Kubo}, \citenamefont {Kusaka}, \citenamefont
  {Motobayashi}, \citenamefont {Nakamura}, \citenamefont {Ohnishi},
  \citenamefont {Okuno}, \citenamefont {Otsu}, \citenamefont {Sako},
  \citenamefont {Sato}, \citenamefont {Shimizu}, \citenamefont {Sekiguchi},
  \citenamefont {Takahashi}, \citenamefont {Tanaka},\ and\ \citenamefont
  {Yoneda}}]{Kobayashi13}%
  \BibitemOpen
  \bibfield  {author} {\bibinfo {author} {\bibfnamefont {T.}~\bibnamefont
  {Kobayashi}}, \bibinfo {author} {\bibfnamefont {N.}~\bibnamefont {Chiga}},
  \bibinfo {author} {\bibfnamefont {T.}~\bibnamefont {Isobe}}, \bibinfo
  {author} {\bibfnamefont {Y.}~\bibnamefont {Kondo}}, \bibinfo {author}
  {\bibfnamefont {T.}~\bibnamefont {Kubo}}, \bibinfo {author} {\bibfnamefont
  {K.}~\bibnamefont {Kusaka}}, \bibinfo {author} {\bibfnamefont
  {T.}~\bibnamefont {Motobayashi}}, \bibinfo {author} {\bibfnamefont
  {T.}~\bibnamefont {Nakamura}}, \bibinfo {author} {\bibfnamefont
  {J.}~\bibnamefont {Ohnishi}}, \bibinfo {author} {\bibfnamefont
  {H.}~\bibnamefont {Okuno}}, \bibinfo {author} {\bibfnamefont
  {H.}~\bibnamefont {Otsu}}, \bibinfo {author} {\bibfnamefont {T.}~\bibnamefont
  {Sako}}, \bibinfo {author} {\bibfnamefont {H.}~\bibnamefont {Sato}}, \bibinfo
  {author} {\bibfnamefont {Y.}~\bibnamefont {Shimizu}}, \bibinfo {author}
  {\bibfnamefont {K.}~\bibnamefont {Sekiguchi}}, \bibinfo {author}
  {\bibfnamefont {K.}~\bibnamefont {Takahashi}}, \bibinfo {author}
  {\bibfnamefont {R.}~\bibnamefont {Tanaka}},\ and\ \bibinfo {author}
  {\bibfnamefont {K.}~\bibnamefont {Yoneda}},\ }\bibfield  {title} {\bibinfo
  {title} {Samurai spectrometer for ri beam experiments},\ }\href
  {https://doi.org/https://doi.org/10.1016/j.nimb.2013.05.089} {\bibfield
  {journal} {\bibinfo  {journal} {Nuclear Instruments and Methods in Physics
  Research Section B: Beam Interactions with Materials and Atoms}\ }\textbf
  {\bibinfo {volume} {317}},\ \bibinfo {pages} {294 } (\bibinfo {year}
  {2013})},\ \bibinfo {note} {xVIth International Conference on ElectroMagnetic
  Isotope Separators and Techniques Related to their Applications, December
  2–7, 2012 at Matsue, Japan}\BibitemShut {NoStop}%
\bibitem [{\citenamefont {Tsang}\ \emph {et~al.}(2020)\citenamefont {Tsang},
  \citenamefont {Estee}, \citenamefont {Wang}, \citenamefont {Barney},
  \citenamefont {Jhang}, \citenamefont {Lynch}, \citenamefont {Zhang},
  \citenamefont {Cerizza}, \citenamefont {Isobe}, \citenamefont {Kaneko},
  \citenamefont {Kurata-Nishimura}, \citenamefont {Lee}, \citenamefont
  {Murakami},\ and\ \citenamefont {Tsang}}]{Tsangc20}%
  \BibitemOpen
  \bibfield  {author} {\bibinfo {author} {\bibfnamefont {C.}~\bibnamefont
  {Tsang}}, \bibinfo {author} {\bibfnamefont {J.}~\bibnamefont {Estee}},
  \bibinfo {author} {\bibfnamefont {R.}~\bibnamefont {Wang}}, \bibinfo {author}
  {\bibfnamefont {J.}~\bibnamefont {Barney}}, \bibinfo {author} {\bibfnamefont
  {G.}~\bibnamefont {Jhang}}, \bibinfo {author} {\bibfnamefont
  {W.}~\bibnamefont {Lynch}}, \bibinfo {author} {\bibfnamefont
  {Z.}~\bibnamefont {Zhang}}, \bibinfo {author} {\bibfnamefont
  {G.}~\bibnamefont {Cerizza}}, \bibinfo {author} {\bibfnamefont
  {T.}~\bibnamefont {Isobe}}, \bibinfo {author} {\bibfnamefont
  {M.}~\bibnamefont {Kaneko}}, \bibinfo {author} {\bibfnamefont
  {M.}~\bibnamefont {Kurata-Nishimura}}, \bibinfo {author} {\bibfnamefont
  {J.}~\bibnamefont {Lee}}, \bibinfo {author} {\bibfnamefont {T.}~\bibnamefont
  {Murakami}},\ and\ \bibinfo {author} {\bibfnamefont {M.}~\bibnamefont
  {Tsang}},\ }\bibfield  {title} {\bibinfo {title} {{Space charge effects in
  the S$\pi$RIT time projection chamber}},\ }\href
  {https://doi.org/https://doi.org/10.1016/j.nima.2020.163477} {\bibfield
  {journal} {\bibinfo  {journal} {Nucl. Instru. Methods in Phys. Res.}\
  }\textbf {\bibinfo {volume} {959}},\ \bibinfo {pages} {163477} (\bibinfo
  {year} {2020})}\BibitemShut {NoStop}%
\bibitem [{\citenamefont {Isobe}\ \emph {et~al.}(2018)\citenamefont {Isobe},
  \citenamefont {Jhang}, \citenamefont {Baba}, \citenamefont {Barney},
  \citenamefont {Baron}, \citenamefont {Cerizza}, \citenamefont {Estee},
  \citenamefont {Kaneko}, \citenamefont {Kurata-Nishimura}, \citenamefont
  {Lee}, \citenamefont {Lynch}, \citenamefont {Murakami}, \citenamefont
  {Nakatsuka}, \citenamefont {Pollacco}, \citenamefont {Powell}, \citenamefont
  {Sakurai}, \citenamefont {Santamaria}, \citenamefont {Suzuki}, \citenamefont
  {Tangwancharoen},\ and\ \citenamefont {Tsang}}]{Iso18}%
  \BibitemOpen
  \bibfield  {author} {\bibinfo {author} {\bibfnamefont {T.}~\bibnamefont
  {Isobe}}, \bibinfo {author} {\bibfnamefont {G.}~\bibnamefont {Jhang}},
  \bibinfo {author} {\bibfnamefont {H.}~\bibnamefont {Baba}}, \bibinfo {author}
  {\bibfnamefont {J.}~\bibnamefont {Barney}}, \bibinfo {author} {\bibfnamefont
  {P.}~\bibnamefont {Baron}}, \bibinfo {author} {\bibfnamefont
  {G.}~\bibnamefont {Cerizza}}, \bibinfo {author} {\bibfnamefont
  {J.}~\bibnamefont {Estee}}, \bibinfo {author} {\bibfnamefont
  {M.}~\bibnamefont {Kaneko}}, \bibinfo {author} {\bibfnamefont
  {M.}~\bibnamefont {Kurata-Nishimura}}, \bibinfo {author} {\bibfnamefont
  {J.}~\bibnamefont {Lee}}, \bibinfo {author} {\bibfnamefont {W.}~\bibnamefont
  {Lynch}}, \bibinfo {author} {\bibfnamefont {T.}~\bibnamefont {Murakami}},
  \bibinfo {author} {\bibfnamefont {N.}~\bibnamefont {Nakatsuka}}, \bibinfo
  {author} {\bibfnamefont {E.}~\bibnamefont {Pollacco}}, \bibinfo {author}
  {\bibfnamefont {W.}~\bibnamefont {Powell}}, \bibinfo {author} {\bibfnamefont
  {H.}~\bibnamefont {Sakurai}}, \bibinfo {author} {\bibfnamefont
  {C.}~\bibnamefont {Santamaria}}, \bibinfo {author} {\bibfnamefont
  {D.}~\bibnamefont {Suzuki}}, \bibinfo {author} {\bibfnamefont
  {S.}~\bibnamefont {Tangwancharoen}},\ and\ \bibinfo {author} {\bibfnamefont
  {M.}~\bibnamefont {Tsang}},\ }\bibfield  {title} {\bibinfo {title}
  {{Application of the Generic Electronics for Time Projection Chamber (GET)
  readout system for heavy Radioactive isotope collision experiments}},\ }\href
  {https://doi.org/https://doi.org/10.1016/j.nima.2018.05.022} {\bibfield
  {journal} {\bibinfo  {journal} {Nucl. Instru. Methods in Phys. Res.}\
  }\textbf {\bibinfo {volume} {899}},\ \bibinfo {pages} {43} (\bibinfo {year}
  {2018})}\BibitemShut {NoStop}%
\bibitem [{\citenamefont {Lee}\ \emph {et~al.}(2020)\citenamefont {Lee},
  \citenamefont {Jhang}, \citenamefont {Cerizza}, \citenamefont {Barney},
  \citenamefont {Estee}, \citenamefont {Isobe}, \citenamefont {Kaneko},
  \citenamefont {Kurata-Nishimura}, \citenamefont {Lynch}, \citenamefont
  {Murakami}, \citenamefont {Tsang}, \citenamefont {Tsang}, \citenamefont
  {Wang}, \citenamefont {Hong}, \citenamefont {McIntosh}, \citenamefont
  {Sakurai}, \citenamefont {Santamaria}, \citenamefont {Shane}, \citenamefont
  {Tangwancharoen}, \citenamefont {Yennello},\ and\ \citenamefont
  {Zhang}}]{Lee20}%
  \BibitemOpen
  \bibfield  {author} {\bibinfo {author} {\bibfnamefont {J.}~\bibnamefont
  {Lee}}, \bibinfo {author} {\bibfnamefont {G.}~\bibnamefont {Jhang}}, \bibinfo
  {author} {\bibfnamefont {G.}~\bibnamefont {Cerizza}}, \bibinfo {author}
  {\bibfnamefont {J.}~\bibnamefont {Barney}}, \bibinfo {author} {\bibfnamefont
  {J.}~\bibnamefont {Estee}}, \bibinfo {author} {\bibfnamefont
  {T.}~\bibnamefont {Isobe}}, \bibinfo {author} {\bibfnamefont
  {M.}~\bibnamefont {Kaneko}}, \bibinfo {author} {\bibfnamefont
  {M.}~\bibnamefont {Kurata-Nishimura}}, \bibinfo {author} {\bibfnamefont
  {W.}~\bibnamefont {Lynch}}, \bibinfo {author} {\bibfnamefont
  {T.}~\bibnamefont {Murakami}}, \bibinfo {author} {\bibfnamefont
  {C.}~\bibnamefont {Tsang}}, \bibinfo {author} {\bibfnamefont
  {M.}~\bibnamefont {Tsang}}, \bibinfo {author} {\bibfnamefont
  {R.}~\bibnamefont {Wang}}, \bibinfo {author} {\bibfnamefont {B.}~\bibnamefont
  {Hong}}, \bibinfo {author} {\bibfnamefont {A.}~\bibnamefont {McIntosh}},
  \bibinfo {author} {\bibfnamefont {H.}~\bibnamefont {Sakurai}}, \bibinfo
  {author} {\bibfnamefont {C.}~\bibnamefont {Santamaria}}, \bibinfo {author}
  {\bibfnamefont {R.}~\bibnamefont {Shane}}, \bibinfo {author} {\bibfnamefont
  {S.}~\bibnamefont {Tangwancharoen}}, \bibinfo {author} {\bibfnamefont
  {S.}~\bibnamefont {Yennello}},\ and\ \bibinfo {author} {\bibfnamefont
  {Y.}~\bibnamefont {Zhang}},\ }\bibfield  {title} {\bibinfo {title} {{Charged
  particle track reconstruction with S$\pi$RIT Time Projection Chamber}},\
  }\href {https://doi.org/https://doi.org/10.1016/j.nima.2020.163840}
  {\bibfield  {journal} {\bibinfo  {journal} {Nucl. Instru. Methods in Phys.
  Res.}\ }\textbf {\bibinfo {volume} {965}},\ \bibinfo {pages} {163840}
  (\bibinfo {year} {2020})}\BibitemShut {NoStop}%
\bibitem [{\citenamefont {Lasko}\ \emph {et~al.}(2017)\citenamefont {Lasko},
  \citenamefont {Adamczyk}, \citenamefont {Brzychczyk}, \citenamefont {Hirnyk},
  \citenamefont {Łukasik}, \citenamefont {Pawłowski}, \citenamefont
  {Pelczar}, \citenamefont {Snoch}, \citenamefont {Sochocka}, \citenamefont
  {Sosin}, \citenamefont {Barney}, \citenamefont {Cerizza}, \citenamefont
  {Estee}, \citenamefont {Isobe}, \citenamefont {Jhang}, \citenamefont
  {Kaneko}, \citenamefont {Kurata-Nishimura}, \citenamefont {Lynch},
  \citenamefont {Murakami}, \citenamefont {Santamaria}, \citenamefont {Tsang},\
  and\ \citenamefont {Zhang}}]{Las17}%
  \BibitemOpen
  \bibfield  {author} {\bibinfo {author} {\bibfnamefont {P.}~\bibnamefont
  {Lasko}}, \bibinfo {author} {\bibfnamefont {M.}~\bibnamefont {Adamczyk}},
  \bibinfo {author} {\bibfnamefont {J.}~\bibnamefont {Brzychczyk}}, \bibinfo
  {author} {\bibfnamefont {P.}~\bibnamefont {Hirnyk}}, \bibinfo {author}
  {\bibfnamefont {J.}~\bibnamefont {Łukasik}}, \bibinfo {author}
  {\bibfnamefont {P.}~\bibnamefont {Pawłowski}}, \bibinfo {author}
  {\bibfnamefont {K.}~\bibnamefont {Pelczar}}, \bibinfo {author} {\bibfnamefont
  {A.}~\bibnamefont {Snoch}}, \bibinfo {author} {\bibfnamefont
  {A.}~\bibnamefont {Sochocka}}, \bibinfo {author} {\bibfnamefont
  {Z.}~\bibnamefont {Sosin}}, \bibinfo {author} {\bibfnamefont
  {J.}~\bibnamefont {Barney}}, \bibinfo {author} {\bibfnamefont
  {G.}~\bibnamefont {Cerizza}}, \bibinfo {author} {\bibfnamefont
  {J.}~\bibnamefont {Estee}}, \bibinfo {author} {\bibfnamefont
  {T.}~\bibnamefont {Isobe}}, \bibinfo {author} {\bibfnamefont
  {G.}~\bibnamefont {Jhang}}, \bibinfo {author} {\bibfnamefont
  {M.}~\bibnamefont {Kaneko}}, \bibinfo {author} {\bibfnamefont
  {M.}~\bibnamefont {Kurata-Nishimura}}, \bibinfo {author} {\bibfnamefont
  {W.}~\bibnamefont {Lynch}}, \bibinfo {author} {\bibfnamefont
  {T.}~\bibnamefont {Murakami}}, \bibinfo {author} {\bibfnamefont
  {C.}~\bibnamefont {Santamaria}}, \bibinfo {author} {\bibfnamefont
  {M.}~\bibnamefont {Tsang}},\ and\ \bibinfo {author} {\bibfnamefont
  {Y.}~\bibnamefont {Zhang}},\ }\bibfield  {title} {\bibinfo {title} {{KATANA
  – A charge-sensitive triggering system for the S$\pi$RIT experiment}},\
  }\href {https://doi.org/https://doi.org/10.1016/j.nima.2017.03.006}
  {\bibfield  {journal} {\bibinfo  {journal} {Nucl. Instru. Methods in Phys.
  Res.}\ }\textbf {\bibinfo {volume} {856}},\ \bibinfo {pages} {92} (\bibinfo
  {year} {2017})}\BibitemShut {NoStop}%
\bibitem [{\citenamefont {Shane}\ \emph {et~al.}(2015)\citenamefont {Shane},
  \citenamefont {McIntosh}, \citenamefont {Isobe}, \citenamefont {Lynch},
  \citenamefont {Baba}, \citenamefont {Barney}, \citenamefont {Chajecki},
  \citenamefont {Chartier}, \citenamefont {Estee}, \citenamefont {Famiano},
  \citenamefont {Hong}, \citenamefont {Ieki}, \citenamefont {Jhang},
  \citenamefont {Lemmon}, \citenamefont {Lu}, \citenamefont {Murakami},
  \citenamefont {Nakatsuka}, \citenamefont {Nishimura}, \citenamefont {Olsen},
  \citenamefont {Powell}, \citenamefont {Sakurai}, \citenamefont {Taketani},
  \citenamefont {Tangwancharoen}, \citenamefont {Tsang}, \citenamefont
  {Usukura}, \citenamefont {Wang}, \citenamefont {Yennello},\ and\
  \citenamefont {Yurkon}}]{Shane15}%
  \BibitemOpen
  \bibfield  {author} {\bibinfo {author} {\bibfnamefont {R.}~\bibnamefont
  {Shane}}, \bibinfo {author} {\bibfnamefont {A.}~\bibnamefont {McIntosh}},
  \bibinfo {author} {\bibfnamefont {T.}~\bibnamefont {Isobe}}, \bibinfo
  {author} {\bibfnamefont {W.}~\bibnamefont {Lynch}}, \bibinfo {author}
  {\bibfnamefont {H.}~\bibnamefont {Baba}}, \bibinfo {author} {\bibfnamefont
  {J.}~\bibnamefont {Barney}}, \bibinfo {author} {\bibfnamefont
  {Z.}~\bibnamefont {Chajecki}}, \bibinfo {author} {\bibfnamefont
  {M.}~\bibnamefont {Chartier}}, \bibinfo {author} {\bibfnamefont
  {J.}~\bibnamefont {Estee}}, \bibinfo {author} {\bibfnamefont
  {M.}~\bibnamefont {Famiano}}, \bibinfo {author} {\bibfnamefont
  {B.}~\bibnamefont {Hong}}, \bibinfo {author} {\bibfnamefont {K.}~\bibnamefont
  {Ieki}}, \bibinfo {author} {\bibfnamefont {G.}~\bibnamefont {Jhang}},
  \bibinfo {author} {\bibfnamefont {R.}~\bibnamefont {Lemmon}}, \bibinfo
  {author} {\bibfnamefont {F.}~\bibnamefont {Lu}}, \bibinfo {author}
  {\bibfnamefont {T.}~\bibnamefont {Murakami}}, \bibinfo {author}
  {\bibfnamefont {N.}~\bibnamefont {Nakatsuka}}, \bibinfo {author}
  {\bibfnamefont {M.}~\bibnamefont {Nishimura}}, \bibinfo {author}
  {\bibfnamefont {R.}~\bibnamefont {Olsen}}, \bibinfo {author} {\bibfnamefont
  {W.}~\bibnamefont {Powell}}, \bibinfo {author} {\bibfnamefont
  {H.}~\bibnamefont {Sakurai}}, \bibinfo {author} {\bibfnamefont
  {A.}~\bibnamefont {Taketani}}, \bibinfo {author} {\bibfnamefont
  {S.}~\bibnamefont {Tangwancharoen}}, \bibinfo {author} {\bibfnamefont
  {M.}~\bibnamefont {Tsang}}, \bibinfo {author} {\bibfnamefont
  {T.}~\bibnamefont {Usukura}}, \bibinfo {author} {\bibfnamefont
  {R.}~\bibnamefont {Wang}}, \bibinfo {author} {\bibfnamefont {S.}~\bibnamefont
  {Yennello}},\ and\ \bibinfo {author} {\bibfnamefont {J.}~\bibnamefont
  {Yurkon}},\ }\bibfield  {title} {\bibinfo {title} {S$\pi$rit: A
  time-projection chamber for symmetry-energy studies},\ }\href
  {https://doi.org/https://doi.org/10.1016/j.nima.2015.01.026} {\bibfield
  {journal} {\bibinfo  {journal} {Nucl. Instru. Methods in Phys. Res.}\
  }\textbf {\bibinfo {volume} {784}},\ \bibinfo {pages} {513 } (\bibinfo {year}
  {2015})},\ \bibinfo {note} {symposium on Radiation Measurements and
  Applications 2014 (SORMA XV)}\BibitemShut {NoStop}%
\bibitem [{\citenamefont {{Barney}}(2019)}]{Barney19}%
  \BibitemOpen
  \bibfield  {author} {\bibinfo {author} {\bibfnamefont {J.~E.}\ \bibnamefont
  {{Barney}}},\ }\emph {\bibinfo {title} {{Charged Pion Emission from
  $^{112}$Sn + $^{124}$Sn and $^{124}$Sn + $^{112}$Sn Reactions with the
  S{\ensuremath{\pi}}RIT Time Projection Chamber}}},\ \href@noop {} {Ph.D.
  thesis},\ \bibinfo  {school} {Michigan State University} (\bibinfo {year}
  {2019})\BibitemShut {NoStop}%
\bibitem [{\citenamefont {Reisdorf}\ \emph {et~al.}(2010)\citenamefont
  {Reisdorf}, \citenamefont {Andronic}, \citenamefont {Averbeck}, \citenamefont
  {Benabderrahmane}, \citenamefont {Hartmann}, \citenamefont {Herrmann},
  \citenamefont {Hildenbrand}, \citenamefont {Kang}, \citenamefont {Kim},
  \citenamefont {Kiš}, \citenamefont {Koczoń}, \citenamefont {Kress},
  \citenamefont {Leifels}, \citenamefont {Merschmeyer}, \citenamefont
  {Piasecki}, \citenamefont {Schüttauf}, \citenamefont {Stockmeier},
  \citenamefont {Barret}, \citenamefont {Basrak}, \citenamefont {Bastid},
  \citenamefont {Čaplar}, \citenamefont {Crochet}, \citenamefont {Dupieux},
  \citenamefont {Dželalija}, \citenamefont {Fodor}, \citenamefont {Gasik},
  \citenamefont {Grishkin}, \citenamefont {Hong}, \citenamefont {Kecskemeti},
  \citenamefont {Kirejczyk}, \citenamefont {Korolija}, \citenamefont {Kotte},
  \citenamefont {Lebedev}, \citenamefont {Lopez}, \citenamefont {Matulewicz},
  \citenamefont {Neubert}, \citenamefont {Petrovici}, \citenamefont {Rami},
  \citenamefont {Ryu}, \citenamefont {Seres}, \citenamefont {Sikora},
  \citenamefont {Sim}, \citenamefont {Simion}, \citenamefont
  {Siwek-Wilczyńska}, \citenamefont {Smolyankin}, \citenamefont {Stoicea},
  \citenamefont {Tymiński}, \citenamefont {Wiśniewski}, \citenamefont
  {Wohlfarth}, \citenamefont {Xiao}, \citenamefont {Xu}, \citenamefont
  {Yushmanov},\ and\ \citenamefont {Zhilin}}]{Reisdorf10}%
  \BibitemOpen
  \bibfield  {author} {\bibinfo {author} {\bibfnamefont {W.}~\bibnamefont
  {Reisdorf}}, \bibinfo {author} {\bibfnamefont {A.}~\bibnamefont {Andronic}},
  \bibinfo {author} {\bibfnamefont {R.}~\bibnamefont {Averbeck}}, \bibinfo
  {author} {\bibfnamefont {M.}~\bibnamefont {Benabderrahmane}}, \bibinfo
  {author} {\bibfnamefont {O.}~\bibnamefont {Hartmann}}, \bibinfo {author}
  {\bibfnamefont {N.}~\bibnamefont {Herrmann}}, \bibinfo {author}
  {\bibfnamefont {K.}~\bibnamefont {Hildenbrand}}, \bibinfo {author}
  {\bibfnamefont {T.}~\bibnamefont {Kang}}, \bibinfo {author} {\bibfnamefont
  {Y.}~\bibnamefont {Kim}}, \bibinfo {author} {\bibfnamefont {M.}~\bibnamefont
  {Kiš}}, \bibinfo {author} {\bibfnamefont {P.}~\bibnamefont {Koczoń}},
  \bibinfo {author} {\bibfnamefont {T.}~\bibnamefont {Kress}}, \bibinfo
  {author} {\bibfnamefont {Y.}~\bibnamefont {Leifels}}, \bibinfo {author}
  {\bibfnamefont {M.}~\bibnamefont {Merschmeyer}}, \bibinfo {author}
  {\bibfnamefont {K.}~\bibnamefont {Piasecki}}, \bibinfo {author}
  {\bibfnamefont {A.}~\bibnamefont {Schüttauf}}, \bibinfo {author}
  {\bibfnamefont {M.}~\bibnamefont {Stockmeier}}, \bibinfo {author}
  {\bibfnamefont {V.}~\bibnamefont {Barret}}, \bibinfo {author} {\bibfnamefont
  {Z.}~\bibnamefont {Basrak}}, \bibinfo {author} {\bibfnamefont
  {N.}~\bibnamefont {Bastid}}, \bibinfo {author} {\bibfnamefont
  {R.}~\bibnamefont {Čaplar}}, \bibinfo {author} {\bibfnamefont
  {P.}~\bibnamefont {Crochet}}, \bibinfo {author} {\bibfnamefont
  {P.}~\bibnamefont {Dupieux}}, \bibinfo {author} {\bibfnamefont
  {M.}~\bibnamefont {Dželalija}}, \bibinfo {author} {\bibfnamefont
  {Z.}~\bibnamefont {Fodor}}, \bibinfo {author} {\bibfnamefont
  {P.}~\bibnamefont {Gasik}}, \bibinfo {author} {\bibfnamefont
  {Y.}~\bibnamefont {Grishkin}}, \bibinfo {author} {\bibfnamefont
  {B.}~\bibnamefont {Hong}}, \bibinfo {author} {\bibfnamefont {J.}~\bibnamefont
  {Kecskemeti}}, \bibinfo {author} {\bibfnamefont {M.}~\bibnamefont
  {Kirejczyk}}, \bibinfo {author} {\bibfnamefont {M.}~\bibnamefont {Korolija}},
  \bibinfo {author} {\bibfnamefont {R.}~\bibnamefont {Kotte}}, \bibinfo
  {author} {\bibfnamefont {A.}~\bibnamefont {Lebedev}}, \bibinfo {author}
  {\bibfnamefont {X.}~\bibnamefont {Lopez}}, \bibinfo {author} {\bibfnamefont
  {T.}~\bibnamefont {Matulewicz}}, \bibinfo {author} {\bibfnamefont
  {W.}~\bibnamefont {Neubert}}, \bibinfo {author} {\bibfnamefont
  {M.}~\bibnamefont {Petrovici}}, \bibinfo {author} {\bibfnamefont
  {F.}~\bibnamefont {Rami}}, \bibinfo {author} {\bibfnamefont {M.}~\bibnamefont
  {Ryu}}, \bibinfo {author} {\bibfnamefont {Z.}~\bibnamefont {Seres}}, \bibinfo
  {author} {\bibfnamefont {B.}~\bibnamefont {Sikora}}, \bibinfo {author}
  {\bibfnamefont {K.}~\bibnamefont {Sim}}, \bibinfo {author} {\bibfnamefont
  {V.}~\bibnamefont {Simion}}, \bibinfo {author} {\bibfnamefont
  {K.}~\bibnamefont {Siwek-Wilczyńska}}, \bibinfo {author} {\bibfnamefont
  {V.}~\bibnamefont {Smolyankin}}, \bibinfo {author} {\bibfnamefont
  {G.}~\bibnamefont {Stoicea}}, \bibinfo {author} {\bibfnamefont
  {Z.}~\bibnamefont {Tymiński}}, \bibinfo {author} {\bibfnamefont
  {K.}~\bibnamefont {Wiśniewski}}, \bibinfo {author} {\bibfnamefont
  {D.}~\bibnamefont {Wohlfarth}}, \bibinfo {author} {\bibfnamefont
  {Z.}~\bibnamefont {Xiao}}, \bibinfo {author} {\bibfnamefont {H.}~\bibnamefont
  {Xu}}, \bibinfo {author} {\bibfnamefont {I.}~\bibnamefont {Yushmanov}},\ and\
  \bibinfo {author} {\bibfnamefont {A.}~\bibnamefont {Zhilin}},\ }\bibfield
  {title} {\bibinfo {title} {{Systematics of central heavy ion collisions in
  the 1 A GeV regime}},\ }\href
  {https://doi.org/https://doi.org/10.1016/j.nuclphysa.2010.09.008} {\bibfield
  {journal} {\bibinfo  {journal} {Nucl. Phys. A}\ }\textbf {\bibinfo {volume}
  {848}},\ \bibinfo {pages} {366 } (\bibinfo {year} {2010})}\BibitemShut
  {NoStop}%
\bibitem [{\citenamefont {Agostinelli}\ \emph {et~al.}(2003)\citenamefont
  {Agostinelli}, \citenamefont {Allison}, \citenamefont {Amako}, \citenamefont
  {Apostolakis}, \citenamefont {Araujo}, \citenamefont {Arce}, \citenamefont
  {Asai}, \citenamefont {Axen}, \citenamefont {Banerjee}, \citenamefont
  {Barrand}, \citenamefont {Behner}, \citenamefont {Bellagamba}, \citenamefont
  {Boudreau}, \citenamefont {Broglia}, \citenamefont {Brunengo}, \citenamefont
  {Burkhardt}, \citenamefont {Chauvie}, \citenamefont {Chuma}, \citenamefont
  {Chytracek}, \citenamefont {Cooperman}, \citenamefont {Cosmo}, \citenamefont
  {Degtyarenko}, \citenamefont {Dell'Acqua}, \citenamefont {Depaola},
  \citenamefont {Dietrich}, \citenamefont {Enami}, \citenamefont {Feliciello},
  \citenamefont {Ferguson}, \citenamefont {Fesefeldt}, \citenamefont {Folger},
  \citenamefont {Foppiano}, \citenamefont {Forti}, \citenamefont {Garelli},
  \citenamefont {Giani}, \citenamefont {Giannitrapani}, \citenamefont {Gibin},
  \citenamefont {{Gómez Cadenas}}, \citenamefont {González}, \citenamefont
  {{Gracia Abril}}, \citenamefont {Greeniaus}, \citenamefont {Greiner},
  \citenamefont {Grichine}, \citenamefont {Grossheim}, \citenamefont
  {Guatelli}, \citenamefont {Gumplinger}, \citenamefont {Hamatsu},
  \citenamefont {Hashimoto}, \citenamefont {Hasui}, \citenamefont {Heikkinen},
  \citenamefont {Howard}, \citenamefont {Ivanchenko}, \citenamefont {Johnson},
  \citenamefont {Jones}, \citenamefont {Kallenbach}, \citenamefont {Kanaya},
  \citenamefont {Kawabata}, \citenamefont {Kawabata}, \citenamefont {Kawaguti},
  \citenamefont {Kelner}, \citenamefont {Kent}, \citenamefont {Kimura},
  \citenamefont {Kodama}, \citenamefont {Kokoulin}, \citenamefont {Kossov},
  \citenamefont {Kurashige}, \citenamefont {Lamanna}, \citenamefont {Lampén},
  \citenamefont {Lara}, \citenamefont {Lefebure}, \citenamefont {Lei},
  \citenamefont {Liendl}, \citenamefont {Lockman}, \citenamefont {Longo},
  \citenamefont {Magni}, \citenamefont {Maire}, \citenamefont {Medernach},
  \citenamefont {Minamimoto}, \citenamefont {{Mora de Freitas}}, \citenamefont
  {Morita}, \citenamefont {Murakami}, \citenamefont {Nagamatu}, \citenamefont
  {Nartallo}, \citenamefont {Nieminen}, \citenamefont {Nishimura},
  \citenamefont {Ohtsubo}, \citenamefont {Okamura}, \citenamefont {O'Neale},
  \citenamefont {Oohata}, \citenamefont {Paech}, \citenamefont {Perl},
  \citenamefont {Pfeiffer}, \citenamefont {Pia}, \citenamefont {Ranjard},
  \citenamefont {Rybin}, \citenamefont {Sadilov}, \citenamefont {{Di Salvo}},
  \citenamefont {Santin}, \citenamefont {Sasaki}, \citenamefont {Savvas},
  \citenamefont {Sawada}, \citenamefont {Scherer}, \citenamefont {Sei},
  \citenamefont {Sirotenko}, \citenamefont {Smith}, \citenamefont {Starkov},
  \citenamefont {Stoecker}, \citenamefont {Sulkimo}, \citenamefont {Takahata},
  \citenamefont {Tanaka}, \citenamefont {Tcherniaev}, \citenamefont {{Safai
  Tehrani}}, \citenamefont {Tropeano}, \citenamefont {Truscott}, \citenamefont
  {Uno}, \citenamefont {Urban}, \citenamefont {Urban}, \citenamefont {Verderi},
  \citenamefont {Walkden}, \citenamefont {Wander}, \citenamefont {Weber},
  \citenamefont {Wellisch}, \citenamefont {Wenaus}, \citenamefont {Williams},
  \citenamefont {Wright}, \citenamefont {Yamada}, \citenamefont {Yoshida},\
  and\ \citenamefont {Zschiesche}}]{Agostinelli03}%
  \BibitemOpen
  \bibfield  {author} {\bibinfo {author} {\bibfnamefont {S.}~\bibnamefont
  {Agostinelli}}, \bibinfo {author} {\bibfnamefont {J.}~\bibnamefont
  {Allison}}, \bibinfo {author} {\bibfnamefont {K.}~\bibnamefont {Amako}},
  \bibinfo {author} {\bibfnamefont {J.}~\bibnamefont {Apostolakis}}, \bibinfo
  {author} {\bibfnamefont {H.}~\bibnamefont {Araujo}}, \bibinfo {author}
  {\bibfnamefont {P.}~\bibnamefont {Arce}}, \bibinfo {author} {\bibfnamefont
  {M.}~\bibnamefont {Asai}}, \bibinfo {author} {\bibfnamefont {D.}~\bibnamefont
  {Axen}}, \bibinfo {author} {\bibfnamefont {S.}~\bibnamefont {Banerjee}},
  \bibinfo {author} {\bibfnamefont {G.}~\bibnamefont {Barrand}}, \bibinfo
  {author} {\bibfnamefont {F.}~\bibnamefont {Behner}}, \bibinfo {author}
  {\bibfnamefont {L.}~\bibnamefont {Bellagamba}}, \bibinfo {author}
  {\bibfnamefont {J.}~\bibnamefont {Boudreau}}, \bibinfo {author}
  {\bibfnamefont {L.}~\bibnamefont {Broglia}}, \bibinfo {author} {\bibfnamefont
  {A.}~\bibnamefont {Brunengo}}, \bibinfo {author} {\bibfnamefont
  {H.}~\bibnamefont {Burkhardt}}, \bibinfo {author} {\bibfnamefont
  {S.}~\bibnamefont {Chauvie}}, \bibinfo {author} {\bibfnamefont
  {J.}~\bibnamefont {Chuma}}, \bibinfo {author} {\bibfnamefont
  {R.}~\bibnamefont {Chytracek}}, \bibinfo {author} {\bibfnamefont
  {G.}~\bibnamefont {Cooperman}}, \bibinfo {author} {\bibfnamefont
  {G.}~\bibnamefont {Cosmo}}, \bibinfo {author} {\bibfnamefont
  {P.}~\bibnamefont {Degtyarenko}}, \bibinfo {author} {\bibfnamefont
  {A.}~\bibnamefont {Dell'Acqua}}, \bibinfo {author} {\bibfnamefont
  {G.}~\bibnamefont {Depaola}}, \bibinfo {author} {\bibfnamefont
  {D.}~\bibnamefont {Dietrich}}, \bibinfo {author} {\bibfnamefont
  {R.}~\bibnamefont {Enami}}, \bibinfo {author} {\bibfnamefont
  {A.}~\bibnamefont {Feliciello}}, \bibinfo {author} {\bibfnamefont
  {C.}~\bibnamefont {Ferguson}}, \bibinfo {author} {\bibfnamefont
  {H.}~\bibnamefont {Fesefeldt}}, \bibinfo {author} {\bibfnamefont
  {G.}~\bibnamefont {Folger}}, \bibinfo {author} {\bibfnamefont
  {F.}~\bibnamefont {Foppiano}}, \bibinfo {author} {\bibfnamefont
  {A.}~\bibnamefont {Forti}}, \bibinfo {author} {\bibfnamefont
  {S.}~\bibnamefont {Garelli}}, \bibinfo {author} {\bibfnamefont
  {S.}~\bibnamefont {Giani}}, \bibinfo {author} {\bibfnamefont
  {R.}~\bibnamefont {Giannitrapani}}, \bibinfo {author} {\bibfnamefont
  {D.}~\bibnamefont {Gibin}}, \bibinfo {author} {\bibfnamefont
  {J.}~\bibnamefont {{Gómez Cadenas}}}, \bibinfo {author} {\bibfnamefont
  {I.}~\bibnamefont {González}}, \bibinfo {author} {\bibfnamefont
  {G.}~\bibnamefont {{Gracia Abril}}}, \bibinfo {author} {\bibfnamefont
  {G.}~\bibnamefont {Greeniaus}}, \bibinfo {author} {\bibfnamefont
  {W.}~\bibnamefont {Greiner}}, \bibinfo {author} {\bibfnamefont
  {V.}~\bibnamefont {Grichine}}, \bibinfo {author} {\bibfnamefont
  {A.}~\bibnamefont {Grossheim}}, \bibinfo {author} {\bibfnamefont
  {S.}~\bibnamefont {Guatelli}}, \bibinfo {author} {\bibfnamefont
  {P.}~\bibnamefont {Gumplinger}}, \bibinfo {author} {\bibfnamefont
  {R.}~\bibnamefont {Hamatsu}}, \bibinfo {author} {\bibfnamefont
  {K.}~\bibnamefont {Hashimoto}}, \bibinfo {author} {\bibfnamefont
  {H.}~\bibnamefont {Hasui}}, \bibinfo {author} {\bibfnamefont
  {A.}~\bibnamefont {Heikkinen}}, \bibinfo {author} {\bibfnamefont
  {A.}~\bibnamefont {Howard}}, \bibinfo {author} {\bibfnamefont
  {V.}~\bibnamefont {Ivanchenko}}, \bibinfo {author} {\bibfnamefont
  {A.}~\bibnamefont {Johnson}}, \bibinfo {author} {\bibfnamefont
  {F.}~\bibnamefont {Jones}}, \bibinfo {author} {\bibfnamefont
  {J.}~\bibnamefont {Kallenbach}}, \bibinfo {author} {\bibfnamefont
  {N.}~\bibnamefont {Kanaya}}, \bibinfo {author} {\bibfnamefont
  {M.}~\bibnamefont {Kawabata}}, \bibinfo {author} {\bibfnamefont
  {Y.}~\bibnamefont {Kawabata}}, \bibinfo {author} {\bibfnamefont
  {M.}~\bibnamefont {Kawaguti}}, \bibinfo {author} {\bibfnamefont
  {S.}~\bibnamefont {Kelner}}, \bibinfo {author} {\bibfnamefont
  {P.}~\bibnamefont {Kent}}, \bibinfo {author} {\bibfnamefont {A.}~\bibnamefont
  {Kimura}}, \bibinfo {author} {\bibfnamefont {T.}~\bibnamefont {Kodama}},
  \bibinfo {author} {\bibfnamefont {R.}~\bibnamefont {Kokoulin}}, \bibinfo
  {author} {\bibfnamefont {M.}~\bibnamefont {Kossov}}, \bibinfo {author}
  {\bibfnamefont {H.}~\bibnamefont {Kurashige}}, \bibinfo {author}
  {\bibfnamefont {E.}~\bibnamefont {Lamanna}}, \bibinfo {author} {\bibfnamefont
  {T.}~\bibnamefont {Lampén}}, \bibinfo {author} {\bibfnamefont
  {V.}~\bibnamefont {Lara}}, \bibinfo {author} {\bibfnamefont {V.}~\bibnamefont
  {Lefebure}}, \bibinfo {author} {\bibfnamefont {F.}~\bibnamefont {Lei}},
  \bibinfo {author} {\bibfnamefont {M.}~\bibnamefont {Liendl}}, \bibinfo
  {author} {\bibfnamefont {W.}~\bibnamefont {Lockman}}, \bibinfo {author}
  {\bibfnamefont {F.}~\bibnamefont {Longo}}, \bibinfo {author} {\bibfnamefont
  {S.}~\bibnamefont {Magni}}, \bibinfo {author} {\bibfnamefont
  {M.}~\bibnamefont {Maire}}, \bibinfo {author} {\bibfnamefont
  {E.}~\bibnamefont {Medernach}}, \bibinfo {author} {\bibfnamefont
  {K.}~\bibnamefont {Minamimoto}}, \bibinfo {author} {\bibfnamefont
  {P.}~\bibnamefont {{Mora de Freitas}}}, \bibinfo {author} {\bibfnamefont
  {Y.}~\bibnamefont {Morita}}, \bibinfo {author} {\bibfnamefont
  {K.}~\bibnamefont {Murakami}}, \bibinfo {author} {\bibfnamefont
  {M.}~\bibnamefont {Nagamatu}}, \bibinfo {author} {\bibfnamefont
  {R.}~\bibnamefont {Nartallo}}, \bibinfo {author} {\bibfnamefont
  {P.}~\bibnamefont {Nieminen}}, \bibinfo {author} {\bibfnamefont
  {T.}~\bibnamefont {Nishimura}}, \bibinfo {author} {\bibfnamefont
  {K.}~\bibnamefont {Ohtsubo}}, \bibinfo {author} {\bibfnamefont
  {M.}~\bibnamefont {Okamura}}, \bibinfo {author} {\bibfnamefont
  {S.}~\bibnamefont {O'Neale}}, \bibinfo {author} {\bibfnamefont
  {Y.}~\bibnamefont {Oohata}}, \bibinfo {author} {\bibfnamefont
  {K.}~\bibnamefont {Paech}}, \bibinfo {author} {\bibfnamefont
  {J.}~\bibnamefont {Perl}}, \bibinfo {author} {\bibfnamefont {A.}~\bibnamefont
  {Pfeiffer}}, \bibinfo {author} {\bibfnamefont {M.}~\bibnamefont {Pia}},
  \bibinfo {author} {\bibfnamefont {F.}~\bibnamefont {Ranjard}}, \bibinfo
  {author} {\bibfnamefont {A.}~\bibnamefont {Rybin}}, \bibinfo {author}
  {\bibfnamefont {S.}~\bibnamefont {Sadilov}}, \bibinfo {author} {\bibfnamefont
  {E.}~\bibnamefont {{Di Salvo}}}, \bibinfo {author} {\bibfnamefont
  {G.}~\bibnamefont {Santin}}, \bibinfo {author} {\bibfnamefont
  {T.}~\bibnamefont {Sasaki}}, \bibinfo {author} {\bibfnamefont
  {N.}~\bibnamefont {Savvas}}, \bibinfo {author} {\bibfnamefont
  {Y.}~\bibnamefont {Sawada}}, \bibinfo {author} {\bibfnamefont
  {S.}~\bibnamefont {Scherer}}, \bibinfo {author} {\bibfnamefont
  {S.}~\bibnamefont {Sei}}, \bibinfo {author} {\bibfnamefont {V.}~\bibnamefont
  {Sirotenko}}, \bibinfo {author} {\bibfnamefont {D.}~\bibnamefont {Smith}},
  \bibinfo {author} {\bibfnamefont {N.}~\bibnamefont {Starkov}}, \bibinfo
  {author} {\bibfnamefont {H.}~\bibnamefont {Stoecker}}, \bibinfo {author}
  {\bibfnamefont {J.}~\bibnamefont {Sulkimo}}, \bibinfo {author} {\bibfnamefont
  {M.}~\bibnamefont {Takahata}}, \bibinfo {author} {\bibfnamefont
  {S.}~\bibnamefont {Tanaka}}, \bibinfo {author} {\bibfnamefont
  {E.}~\bibnamefont {Tcherniaev}}, \bibinfo {author} {\bibfnamefont
  {E.}~\bibnamefont {{Safai Tehrani}}}, \bibinfo {author} {\bibfnamefont
  {M.}~\bibnamefont {Tropeano}}, \bibinfo {author} {\bibfnamefont
  {P.}~\bibnamefont {Truscott}}, \bibinfo {author} {\bibfnamefont
  {H.}~\bibnamefont {Uno}}, \bibinfo {author} {\bibfnamefont {L.}~\bibnamefont
  {Urban}}, \bibinfo {author} {\bibfnamefont {P.}~\bibnamefont {Urban}},
  \bibinfo {author} {\bibfnamefont {M.}~\bibnamefont {Verderi}}, \bibinfo
  {author} {\bibfnamefont {A.}~\bibnamefont {Walkden}}, \bibinfo {author}
  {\bibfnamefont {W.}~\bibnamefont {Wander}}, \bibinfo {author} {\bibfnamefont
  {H.}~\bibnamefont {Weber}}, \bibinfo {author} {\bibfnamefont
  {J.}~\bibnamefont {Wellisch}}, \bibinfo {author} {\bibfnamefont
  {T.}~\bibnamefont {Wenaus}}, \bibinfo {author} {\bibfnamefont
  {D.}~\bibnamefont {Williams}}, \bibinfo {author} {\bibfnamefont
  {D.}~\bibnamefont {Wright}}, \bibinfo {author} {\bibfnamefont
  {T.}~\bibnamefont {Yamada}}, \bibinfo {author} {\bibfnamefont
  {H.}~\bibnamefont {Yoshida}},\ and\ \bibinfo {author} {\bibfnamefont
  {D.}~\bibnamefont {Zschiesche}},\ }\bibfield  {title} {\bibinfo {title}
  {{GEANT4—a simulation toolkit}},\ }\href
  {https://doi.org/https://doi.org/10.1016/S0168-9002(03)01368-8} {\bibfield
  {journal} {\bibinfo  {journal} {Nucl. Instru. Methods in Phys. Res.}\
  }\textbf {\bibinfo {volume} {506}},\ \bibinfo {pages} {250 } (\bibinfo {year}
  {2003})}\BibitemShut {NoStop}%
\bibitem [{\citenamefont {Ikeno}\ \emph {et~al.}(2016)\citenamefont {Ikeno},
  \citenamefont {Ono}, \citenamefont {Nara},\ and\ \citenamefont
  {Ohnishi}}]{Ike16}%
  \BibitemOpen
  \bibfield  {author} {\bibinfo {author} {\bibfnamefont {N.}~\bibnamefont
  {Ikeno}}, \bibinfo {author} {\bibfnamefont {A.}~\bibnamefont {Ono}}, \bibinfo
  {author} {\bibfnamefont {Y.}~\bibnamefont {Nara}},\ and\ \bibinfo {author}
  {\bibfnamefont {A.}~\bibnamefont {Ohnishi}},\ }\bibfield  {title} {\bibinfo
  {title} {Probing neutron-proton dynamics by pions},\ }\href
  {https://doi.org/10.1103/PhysRevC.93.044612} {\bibfield  {journal} {\bibinfo
  {journal} {Phys. Rev. C}\ }\textbf {\bibinfo {volume} {93}},\ \bibinfo
  {pages} {044612} (\bibinfo {year} {2016})}\BibitemShut {NoStop}%
\bibitem [{\citenamefont {Ikeno}\ \emph {et~al.}(2020)\citenamefont {Ikeno},
  \citenamefont {Ono}, \citenamefont {Nara},\ and\ \citenamefont
  {Ohnishi}}]{Ike20}%
  \BibitemOpen
  \bibfield  {author} {\bibinfo {author} {\bibfnamefont {N.}~\bibnamefont
  {Ikeno}}, \bibinfo {author} {\bibfnamefont {A.}~\bibnamefont {Ono}}, \bibinfo
  {author} {\bibfnamefont {Y.}~\bibnamefont {Nara}},\ and\ \bibinfo {author}
  {\bibfnamefont {A.}~\bibnamefont {Ohnishi}},\ }\bibfield  {title} {\bibinfo
  {title} {Effects of pauli blocking on pion production in central collisions
  of neutron-rich nuclei},\ }\href
  {https://doi.org/10.1103/PhysRevC.101.034607} {\bibfield  {journal} {\bibinfo
   {journal} {Phys. Rev. C}\ }\textbf {\bibinfo {volume} {101}},\ \bibinfo
  {pages} {034607} (\bibinfo {year} {2020})}\BibitemShut {NoStop}%
\bibitem [{\citenamefont {Cozma}(2016)}]{Coz16}%
  \BibitemOpen
  \bibfield  {author} {\bibinfo {author} {\bibfnamefont {M.}~\bibnamefont
  {Cozma}},\ }\bibfield  {title} {\bibinfo {title} {{The impact of energy
  conservation in transport models on the $\pi^-$/$\pi^+$ multiplicity ratio in
  heavy-ion collisions and the symmetry energy}},\ }\href
  {https://doi.org/https://doi.org/10.1016/j.physletb.2015.12.015} {\bibfield
  {journal} {\bibinfo  {journal} {Phys. Lett. B}\ }\textbf {\bibinfo {volume}
  {753}},\ \bibinfo {pages} {166} (\bibinfo {year} {2016})}\BibitemShut
  {NoStop}%
\bibitem [{\citenamefont {Cozma}(2017)}]{Coz17}%
  \BibitemOpen
  \bibfield  {author} {\bibinfo {author} {\bibfnamefont {M.~D.}\ \bibnamefont
  {Cozma}},\ }\bibfield  {title} {\bibinfo {title} {{Constraining the density
  dependence of the symmetry energy using the multiplicity and average
  ${p}_{T}$ ratios of charged pions}},\ }\href
  {https://doi.org/10.1103/PhysRevC.95.014601} {\bibfield  {journal} {\bibinfo
  {journal} {Phys. Rev. C}\ }\textbf {\bibinfo {volume} {95}},\ \bibinfo
  {pages} {014601} (\bibinfo {year} {2017})}\BibitemShut {NoStop}%
\bibitem [{\citenamefont {Hartnack}\ \emph {et~al.}(1998)\citenamefont
  {Hartnack}, \citenamefont {Puri}, \citenamefont {Aichelin}, \citenamefont
  {Konopka}, \citenamefont {Bass}, \citenamefont {Stoecker},\ and\
  \citenamefont {Greiner}}]{hartnack1998}%
  \BibitemOpen
  \bibfield  {author} {\bibinfo {author} {\bibfnamefont {C.}~\bibnamefont
  {Hartnack}}, \bibinfo {author} {\bibfnamefont {R.~K.}\ \bibnamefont {Puri}},
  \bibinfo {author} {\bibfnamefont {J.}~\bibnamefont {Aichelin}}, \bibinfo
  {author} {\bibfnamefont {J.}~\bibnamefont {Konopka}}, \bibinfo {author}
  {\bibfnamefont {S.}~\bibnamefont {Bass}}, \bibinfo {author} {\bibfnamefont
  {H.}~\bibnamefont {Stoecker}},\ and\ \bibinfo {author} {\bibfnamefont
  {W.}~\bibnamefont {Greiner}},\ }\bibfield  {title} {\bibinfo {title}
  {Modelling the many-body dynamics of heavy ion collisions: Present status and
  future perspective},\ }\href@noop {} {\bibfield  {journal} {\bibinfo
  {journal} {Euro. Phys. J. A}\ }\textbf {\bibinfo {volume} {1}},\ \bibinfo
  {pages} {151} (\bibinfo {year} {1998})}\BibitemShut {NoStop}%
\bibitem [{\citenamefont {Le~Fevre}\ \emph {et~al.}(2016)\citenamefont
  {Le~Fevre}, \citenamefont {Leifels}, \citenamefont {Reisdorf}, \citenamefont
  {Aichelin},\ and\ \citenamefont {Hartnack}}]{le2016}%
  \BibitemOpen
  \bibfield  {author} {\bibinfo {author} {\bibfnamefont {A.}~\bibnamefont
  {Le~Fevre}}, \bibinfo {author} {\bibfnamefont {Y.}~\bibnamefont {Leifels}},
  \bibinfo {author} {\bibfnamefont {W.}~\bibnamefont {Reisdorf}}, \bibinfo
  {author} {\bibfnamefont {J.}~\bibnamefont {Aichelin}},\ and\ \bibinfo
  {author} {\bibfnamefont {C.}~\bibnamefont {Hartnack}},\ }\bibfield  {title}
  {\bibinfo {title} {Constraining the nuclear matter equation of state around
  twice saturation density},\ }\href@noop {} {\bibfield  {journal} {\bibinfo
  {journal} {Nucl. Phys. A}\ }\textbf {\bibinfo {volume} {945}},\ \bibinfo
  {pages} {112} (\bibinfo {year} {2016})}\BibitemShut {NoStop}%
\bibitem [{\citenamefont {Zhang}\ \emph {et~al.}(2014)\citenamefont {Zhang},
  \citenamefont {Tsang}, \citenamefont {Li},\ and\ \citenamefont
  {Liu}}]{ZHANG14}%
  \BibitemOpen
  \bibfield  {author} {\bibinfo {author} {\bibfnamefont {Y.}~\bibnamefont
  {Zhang}}, \bibinfo {author} {\bibfnamefont {M.}~\bibnamefont {Tsang}},
  \bibinfo {author} {\bibfnamefont {Z.}~\bibnamefont {Li}},\ and\ \bibinfo
  {author} {\bibfnamefont {H.}~\bibnamefont {Liu}},\ }\bibfield  {title}
  {\bibinfo {title} {Constraints on nucleon effective mass splitting with heavy
  ion collisions},\ }\href
  {https://doi.org/https://doi.org/10.1016/j.physletb.2014.03.030} {\bibfield
  {journal} {\bibinfo  {journal} {Phys. Lett. B}\ }\textbf {\bibinfo {volume}
  {732}},\ \bibinfo {pages} {186} (\bibinfo {year} {2014})}\BibitemShut
  {NoStop}%
\bibitem [{\citenamefont {Xu}\ \emph {et~al.}(2016)\citenamefont {Xu},
  \citenamefont {Chen}, \citenamefont {Tsang}, \citenamefont {Wolter},
  \citenamefont {Zhang}, \citenamefont {Aichelin}, \citenamefont {Colonna},
  \citenamefont {Cozma}, \citenamefont {Danielewicz}, \citenamefont {Feng},
  \citenamefont {Le~F\`evre}, \citenamefont {Gaitanos}, \citenamefont
  {Hartnack}, \citenamefont {Kim}, \citenamefont {Kim}, \citenamefont {Ko},
  \citenamefont {Li}, \citenamefont {Li}, \citenamefont {Li}, \citenamefont
  {Napolitani}, \citenamefont {Ono}, \citenamefont {Papa}, \citenamefont
  {Song}, \citenamefont {Su}, \citenamefont {Tian}, \citenamefont {Wang},
  \citenamefont {Wang}, \citenamefont {Weil}, \citenamefont {Xie},
  \citenamefont {Zhang},\ and\ \citenamefont {Zhang}}]{Xu16}%
  \BibitemOpen
  \bibfield  {author} {\bibinfo {author} {\bibfnamefont {J.}~\bibnamefont
  {Xu}}, \bibinfo {author} {\bibfnamefont {L.-W.}\ \bibnamefont {Chen}},
  \bibinfo {author} {\bibfnamefont {M.~B.}\ \bibnamefont {Tsang}}, \bibinfo
  {author} {\bibfnamefont {H.}~\bibnamefont {Wolter}}, \bibinfo {author}
  {\bibfnamefont {Y.-X.}\ \bibnamefont {Zhang}}, \bibinfo {author}
  {\bibfnamefont {J.}~\bibnamefont {Aichelin}}, \bibinfo {author}
  {\bibfnamefont {M.}~\bibnamefont {Colonna}}, \bibinfo {author} {\bibfnamefont
  {D.}~\bibnamefont {Cozma}}, \bibinfo {author} {\bibfnamefont
  {P.}~\bibnamefont {Danielewicz}}, \bibinfo {author} {\bibfnamefont {Z.-Q.}\
  \bibnamefont {Feng}}, \bibinfo {author} {\bibfnamefont {A.}~\bibnamefont
  {Le~F\`evre}}, \bibinfo {author} {\bibfnamefont {T.}~\bibnamefont
  {Gaitanos}}, \bibinfo {author} {\bibfnamefont {C.}~\bibnamefont {Hartnack}},
  \bibinfo {author} {\bibfnamefont {K.}~\bibnamefont {Kim}}, \bibinfo {author}
  {\bibfnamefont {Y.}~\bibnamefont {Kim}}, \bibinfo {author} {\bibfnamefont
  {C.-M.}\ \bibnamefont {Ko}}, \bibinfo {author} {\bibfnamefont {B.-A.}\
  \bibnamefont {Li}}, \bibinfo {author} {\bibfnamefont {Q.-F.}\ \bibnamefont
  {Li}}, \bibinfo {author} {\bibfnamefont {Z.-X.}\ \bibnamefont {Li}}, \bibinfo
  {author} {\bibfnamefont {P.}~\bibnamefont {Napolitani}}, \bibinfo {author}
  {\bibfnamefont {A.}~\bibnamefont {Ono}}, \bibinfo {author} {\bibfnamefont
  {M.}~\bibnamefont {Papa}}, \bibinfo {author} {\bibfnamefont {T.}~\bibnamefont
  {Song}}, \bibinfo {author} {\bibfnamefont {J.}~\bibnamefont {Su}}, \bibinfo
  {author} {\bibfnamefont {J.-L.}\ \bibnamefont {Tian}}, \bibinfo {author}
  {\bibfnamefont {N.}~\bibnamefont {Wang}}, \bibinfo {author} {\bibfnamefont
  {Y.-J.}\ \bibnamefont {Wang}}, \bibinfo {author} {\bibfnamefont
  {J.}~\bibnamefont {Weil}}, \bibinfo {author} {\bibfnamefont {W.-J.}\
  \bibnamefont {Xie}}, \bibinfo {author} {\bibfnamefont {F.-S.}\ \bibnamefont
  {Zhang}},\ and\ \bibinfo {author} {\bibfnamefont {G.-Q.}\ \bibnamefont
  {Zhang}},\ }\bibfield  {title} {\bibinfo {title} {{Understanding transport
  simulations of heavy-ion collisions at $100$ and $400$ A MeV: Comparison of
  heavy-ion transport codes under controlled conditions}},\ }\href
  {https://doi.org/10.1103/PhysRevC.93.044609} {\bibfield  {journal} {\bibinfo
  {journal} {Phys. Rev. C}\ }\textbf {\bibinfo {volume} {93}},\ \bibinfo
  {pages} {044609} (\bibinfo {year} {2016})}\BibitemShut {NoStop}%
\bibitem [{\citenamefont {Zhang}\ \emph {et~al.}(2018)\citenamefont {Zhang},
  \citenamefont {Wang}, \citenamefont {Colonna}, \citenamefont {Danielewicz},
  \citenamefont {Ono}, \citenamefont {Tsang}, \citenamefont {Wolter},
  \citenamefont {Xu}, \citenamefont {Chen}, \citenamefont {Cozma},
  \citenamefont {Feng}, \citenamefont {Das~Gupta}, \citenamefont {Ikeno},
  \citenamefont {Ko}, \citenamefont {Li}, \citenamefont {Li}, \citenamefont
  {Li}, \citenamefont {Mallik}, \citenamefont {Nara}, \citenamefont {Ogawa},
  \citenamefont {Ohnishi}, \citenamefont {Oliinychenko}, \citenamefont {Papa},
  \citenamefont {Petersen}, \citenamefont {Su}, \citenamefont {Song},
  \citenamefont {Weil}, \citenamefont {Wang}, \citenamefont {Zhang},\ and\
  \citenamefont {Zhang}}]{Zhang18}%
  \BibitemOpen
  \bibfield  {author} {\bibinfo {author} {\bibfnamefont {Y.-X.}\ \bibnamefont
  {Zhang}}, \bibinfo {author} {\bibfnamefont {Y.-J.}\ \bibnamefont {Wang}},
  \bibinfo {author} {\bibfnamefont {M.}~\bibnamefont {Colonna}}, \bibinfo
  {author} {\bibfnamefont {P.}~\bibnamefont {Danielewicz}}, \bibinfo {author}
  {\bibfnamefont {A.}~\bibnamefont {Ono}}, \bibinfo {author} {\bibfnamefont
  {M.~B.}\ \bibnamefont {Tsang}}, \bibinfo {author} {\bibfnamefont
  {H.}~\bibnamefont {Wolter}}, \bibinfo {author} {\bibfnamefont
  {J.}~\bibnamefont {Xu}}, \bibinfo {author} {\bibfnamefont {L.-W.}\
  \bibnamefont {Chen}}, \bibinfo {author} {\bibfnamefont {D.}~\bibnamefont
  {Cozma}}, \bibinfo {author} {\bibfnamefont {Z.-Q.}\ \bibnamefont {Feng}},
  \bibinfo {author} {\bibfnamefont {S.}~\bibnamefont {Das~Gupta}}, \bibinfo
  {author} {\bibfnamefont {N.}~\bibnamefont {Ikeno}}, \bibinfo {author}
  {\bibfnamefont {C.-M.}\ \bibnamefont {Ko}}, \bibinfo {author} {\bibfnamefont
  {B.-A.}\ \bibnamefont {Li}}, \bibinfo {author} {\bibfnamefont {Q.-F.}\
  \bibnamefont {Li}}, \bibinfo {author} {\bibfnamefont {Z.-X.}\ \bibnamefont
  {Li}}, \bibinfo {author} {\bibfnamefont {S.}~\bibnamefont {Mallik}}, \bibinfo
  {author} {\bibfnamefont {Y.}~\bibnamefont {Nara}}, \bibinfo {author}
  {\bibfnamefont {T.}~\bibnamefont {Ogawa}}, \bibinfo {author} {\bibfnamefont
  {A.}~\bibnamefont {Ohnishi}}, \bibinfo {author} {\bibfnamefont
  {D.}~\bibnamefont {Oliinychenko}}, \bibinfo {author} {\bibfnamefont
  {M.}~\bibnamefont {Papa}}, \bibinfo {author} {\bibfnamefont {H.}~\bibnamefont
  {Petersen}}, \bibinfo {author} {\bibfnamefont {J.}~\bibnamefont {Su}},
  \bibinfo {author} {\bibfnamefont {T.}~\bibnamefont {Song}}, \bibinfo {author}
  {\bibfnamefont {J.}~\bibnamefont {Weil}}, \bibinfo {author} {\bibfnamefont
  {N.}~\bibnamefont {Wang}}, \bibinfo {author} {\bibfnamefont {F.-S.}\
  \bibnamefont {Zhang}},\ and\ \bibinfo {author} {\bibfnamefont
  {Z.}~\bibnamefont {Zhang}},\ }\bibfield  {title} {\bibinfo {title}
  {Comparison of heavy-ion transport simulations: Collision integral in a
  box},\ }\href {https://doi.org/10.1103/PhysRevC.97.034625} {\bibfield
  {journal} {\bibinfo  {journal} {Phys. Rev. C}\ }\textbf {\bibinfo {volume}
  {97}},\ \bibinfo {pages} {034625} (\bibinfo {year} {2018})}\BibitemShut
  {NoStop}%
\bibitem [{\citenamefont {Poskanzer}\ and\ \citenamefont
  {Voloshin}(1998)}]{poskanzer1998methods}%
  \BibitemOpen
  \bibfield  {author} {\bibinfo {author} {\bibfnamefont {A.~M.}\ \bibnamefont
  {Poskanzer}}\ and\ \bibinfo {author} {\bibfnamefont {S.~A.}\ \bibnamefont
  {Voloshin}},\ }\bibfield  {title} {\bibinfo {title} {Methods for analyzing
  anisotropic flow in relativistic nuclear collisions},\ }\href@noop {}
  {\bibfield  {journal} {\bibinfo  {journal} {Phys. Rev. C}\ }\textbf {\bibinfo
  {volume} {58}},\ \bibinfo {pages} {1671} (\bibinfo {year}
  {1998})}\BibitemShut {NoStop}%
\end{thebibliography}%


